\documentclass[fleqn,usenatbib]{mnras}

\usepackage{newtxtext,newtxmath}

\usepackage[T1]{fontenc}
\usepackage[normalem]{ulem}

\DeclareRobustCommand{\VAN}[3]{#2}
\let\VANthebibliography\thebibliography
\def\thebibliography{\DeclareRobustCommand{\VAN}[3]{##3}\VANthebibliography}

\usepackage{graphicx}	
\usepackage{subfigure}
\usepackage{multirow}
\usepackage{booktabs}

\title[Evolution of Twins]{Detailed Evolutionary Models for Twins in Sight of New Spectral Data: AN Cam, RS Ari, and V455 Aur}

\author[Y\"{u}cel et al.]{
G\"{o}khan Y\"{u}cel,$^{1}$\thanks{E-mail: gokhannyucel@gmail.com}
Volkan Bak{\i}\c{s}$^{1}$
\\
$^{1}$Akdeniz University, Department of Space Sciences and Technologies, Dumlup{\i}nar Blv., Kamp\"{u}s, 07058, Antalya, TR\\
}

\date{Accepted XXX. Received YYY; in original form ZZZ}

\pubyear{2022}

\begin{document}
\defcitealias{imbert1987}{I87}
\defcitealias{imbert2002}{I02}
\defcitealias{Griffin2001}{G01}
\defcitealias{Griffin2013}{G13}
\defcitealias{Southworth2021}{S21a}
\defcitealias{Southworth2021b}{S21b}

\label{firstpage}
\pagerange{\pageref{firstpage}--\pageref{lastpage}}
\maketitle

\begin{abstract}
We present the evolutionary scenarios for three eclipsing twin ($q(M_2/M_1)\sim$1) binary systems using their combined spectroscopic and photometric data. Using accurate \textit{TESS} photometric data, RV measurements, and spectroscopic data enabled us to calculate fundamental parameters, such as mass and radius, better than 2 percent. The temperature of each component and metallicity of the systems have been obtained via high-resolution spectra. According to our spectral analysis, the metallicity values of AN Cam, RS Ari, and V455 Aur are \text{[M/H]}=\,0.00$\pm$0.12, 0.05$\pm$0.08, and -0.07$\pm$0.07, respectively. Using the derived  metallicity for each system, initial orbital parameters and detailed evolutionary status of these three systems are calculated with high precision by using \textsc{mesa}. According to our analysis, both components of AN Cam have passed the terminal age main-sequence, the primary component of RS Ari is in the giant phase while the secondary component has passed the terminal age main-sequence, finally, both components of V455 Aur are still on the main-sequence. The current ages of the three systems AN Cam, RS Ari, and V455 Aur are 3.0, 3.3, and 1.4 Gyrs, respectively, and they will approximately start to transfer mass between components in 400, 250, and 2700 Myrs, respectively.
\end{abstract}




\section{Introduction}

The evolution of a star mainly depends on its mass and chemical composition, where mass plays a major role in evolution. There are several ways to measure the mass of a star, which is explained substantially by \cite{Serenelli}. Masses of stars, which are in binary systems, can be calculated directly via radial velocities of components if the angular inclination of its orbit is known. For visual binaries, calculation of inclination of the orbit is relatively easy but for eclipsing binaries, one needs the light curve (LC) of the system, which also helps not only masses but also to calculate fundamental physical parameters of the components of the system such as the radius, therefore log g, temperature ratios, light contributions, etc. By combining radial velocity (RV) and LC solutions, fundamental parameters of an eclipsing binary can be deduced within 1\% uncertainties \citep[][]{Andersen,Torres}. This feature of eclipsing binaries and its results have made them the focus point in areas such as stellar formation \citep[e.g.][]{Shu}, stellar evolution \citep[e.g.][]{Baraffe,Hurley,Pietrinferni,Bressan}, stellar populations \citep[e.g.][]{Renzini,Duquennoy,Chabrier,Moe}, empirical MLR relation studies \citep[e.g.][]{Demircan,Benedict,Eker} and many more.

On observational time bases, RV measurements can be obtained relatively easily compared to the LC of an eclipsing binary. Along with technological developments, ground-based robotic/automated small telescopes had took the job of observation of variables stars, which relaxed the pressure on relatively big telescopes for observation time requests \citetext{\citealp[ASAS,][]{Pojmanski}; \citealp[OGLE,][]{Udalski}}. Over the last decade, with a focus on exoplanets and needing more precise observations to find more exoplanets, ground-based photometric telescopes have been superseded by space-based photometric telescopes \citetext{\citealp[Kepler]{Borucki}; \citealp[TESS]{TESS}}. Although the main goal of space-based photometric telescopes is to find exoplanets, the way they work is basically the same for building a LC, which is detecting and measuring the brightness change of a star in a certain time interval. Also, since these telescopes are not affected by the day-night cycle of Earth, they can make uninterrupted observations, which provides valuable LC data for eclipsing binaries with long periods \citep[e.g.][]{Prsa2011,Prsa2022}.

Binaries with a mass ratio bigger than 0.95 are described as twin binaries \citep[][]{Tokovinin2000}. Statistical studies on twin binaries have shown that spectral types of twin binaries are predominantly F-G-K \citep[][]{Halb2003,Simon2009}. Although there are studies to explain the reasons for forming equal components, there is still no exact solution for this twin-phenomenon \citep[][]{Bate2000,Bate2002,Lucy2006,Moe,Kounkel2019,Elbadry2019}. Lately, there have been studies that focus on discovering eclipsing twin binary systems based on photometric surveys and space-based photometric telescopes, which have increased the number of known twin-binaries extensively \citep[][]{Zhang2017,Bakis2020,Yucel2022}. \cite{Bate2019} shows that low metallicity produces low mass-ratio binaries with low-mass stellar primaries ($M_1 = 0.1 - 1.5\, \mathrm{M}_\odot$) but noted that this result has been obtained for a limited number of systems. Clearly, there is still in need for more analyzed twin binary systems to understand this phenomenon. Especially, well-determined metallicities and initial orbital conditions of these systems would greatly help scientists to understand the mystery of how twin binaries have been formed.

In this respect, we have analyzed three eclipsing twin binary systems, AN Cam, RS Ari, and V455 Aur, using photometric and RV data in the literature as well as new high resolution spectral data we have obtained. Fundamental parameters of these three systems together with metallicity values have been obtained, and their evolutionary models have been calculated.

The paper is structured as follows. In \S2, we present our target selection criteria and properties of the observational data used. In \S3, calculated fundamental parameters of these systems are presented. In \S4, we present the detailed evolutionary analysis for three systems. Finally, in \S5, we have discussed our overall results.

\section{Targets and Data}

\subsection{The Target Selection}

The systems that we have analyzed were selected specifically from The 9th Catalogue of Spectroscopic Binary Orbits \citep[SB9,][]{Pourbaix2004} under four conditions, which are 1- they must be a twin binary system, 2- they must have similar total mass to see the different evolutionary stages due to their ages,  3- they must have available LC data, and 4- they must be bright enough to obtain its spectrum with our 24-inch telescope and $R=12000$ resolution spectrograph.

AN Camelopardalis is a twin-binary system with a period of 20.9986 days. Its twin feature was discovered by \citet[hereafter I87]{imbert1987} from RV data. Just recently, \citet[hereafter S21a]{Southworth2021} has calculated the radius of its components by using the TESS data. With help of Gaia EDR3 optical/infrared apparent magnitude values, they calculated the effective temperature of the components. They noted that the primary star is less massive than the secondary one. There is no spectroscopic study of the system.

RS Arietis is a twin binary system with a period of 8.80318 days. Its twin feature was discovered by \citet[hereafter I02]{imbert2002} from RV data. There is no prior study on this system.

V455 Aurigae is a triple system. Its eclipsing nature has been revealed by \cite{Hipparcos} but its nature as a twin binary system within a three-body system has been revealed by analyzing RVs of the system, which were obtained by \cite{Griffin2001}. In that analysis, unusual changes in systemic velocity ($V_\gamma$) values were noticed, and the cause of this was expected from a third body in the system. Afterward with additional observation, \citet[hereafter G13]{Griffin2013} has obtained the third body's orbital parameters and indicates that the mass of the third body could be no less than 0.5 $\mathrm{M}_\odot$. Recently, \citet[hereafter S21b]{Southworth2021b} has obtained the radius of the components in the twin-binary system and radius of the third body by using TESS data, and temperatures of bodies in the system by using theoretical spectra with empirical calibrations. There is no spectral study of this system.

\subsection{RV Data}

RV data for AN Cam, RS Ari, and V455 Aur have been taken from \citetalias{imbert1987}, \citetalias{imbert2002}, and \citetalias{Griffin2013}, respectively. Details can be found in those papers, but we briefly give a summary here.

\citetalias{imbert1987} observed AN Cam at Observatoire de Haute-Provence\footnote{\url{http://www.obs-hp.fr/}} with 1-meter Swiss telescope, equipped with Coravel \citep[][]{Baranne1979} instrument and managed to measure 40 and 34 RVs for the primary and secondary components, respectively. They have noted that exposure times had required 10 to 15 minutes due to the presence of two neighboring components causing the cross-correlation peaks to be shallower.

\citetalias{imbert2002} observed RS Ari in years between 1978-1992, and have measured 62 RVs for both components. Besides two measurements, which had been obtained via 1.5-meters Danish telescope of ESO\footnote{\url{https://eso.org/public/teles-instr/lasilla/danish154/coravel/}}, all measurements were obtained using a 1-meter Swiss telescope, which is located at the Observatoire de Haute-Provence.

\citetalias{Griffin2013} observed V455 Aur in two different observation schedules using the 36-inch telescope equipped with a photoelectric RV spectrometer \citep[][]{Griffin1967}, which is located at the Cambridge Observatory, and managed to measure a total of 200 RVs (100 for primary and 100 for secondary) for the system.

\subsection{Photometric Data}

The photometric data used to study three systems were produced by the Transiting Exoplanet Survey Satellite (\textit{TESS}) of NASA. Although, the main object of this satellite is finding the exoplanets, it is also a good source for eclipsing binaries \citep[][]{Prsa2022}. Details of the spacecraft and camera of TESS have been explained extensively by \cite{TESS}. Therefore, we will give a brief information about the photometric data that we have used to analyze these systems.

AN Cam had been observed by the TESS telescope in three sectors, 19, 25, and 26, with long cadence (1800s) observations. We have used Sector 19, which covers observation dates between 2019/11/27 and 2019/12/24 for analysis since both eclipses are well resolved. AN Cam will also be observed in sectors 52, 53, and 59.

RS Ari had been observed by the TESS telescope in four sectors, 18, 42, 43, and 44, with short cadence (120s) and long cadence (1800s) observations. We have used Sector 18, which covers observation dates between 2019/11/02 and 2019/11/27. RS Ari will also be observed in Sector 58. 

V455 Aur had been observed by the TESS telescope in only one sector (20), with short cadence (120s) observations. Sector 20 covers observation dates between 2019/12/24 and 2020/01/21. V455 Aur will also be observed in Sector 60.

\subsection{Spectral Data}

Spectroscopic observations for determining the temperature of components and metallicity of the system were carried out with a 24-inch telescope (UBT60), which is equipped with a Shelyak Instruments eShel Spectrograph (SIeS). It provides spectra between $\lambda$ 4050 and 8160 in 27 \'{e}chelle orders with a resolving power of $R \sim 12000$. More details on CCD and telescope can be found in \citet[]{Bakis2020}. Since RV data are enough to calculate spectroscopic orbits of the systems we analyzed, we observed each system in one observation night in an orbital phase which the spectral lines of both components are clearly visible. Since two of the systems in our list, AN Cam and RS Ari, are relatively faint for our instruments, 6 spectra with exposure times of 1200s have been collected and combined to reach a sufficient S/N ratio. 3 spectra with exposure times of 600s were sufficient for V455 Aur since it is a relatively bright system. The S/N ratio together with other observing information  for the spectral data are given in Table\,\ref{tab:SpecObs}. The spectra were reduced, wavelength calibrated, and continuum normalized using the Image Reduction and Analysis Facility, {\textsc{iraf}}\footnote{{\sc iraf} is provided by National Optical Astronomy Observatories (NOAO) in Tucson, Arizona, USA \citep[][]{Tody}.}. 

\begin{table}
\caption{Log for spectroscopic observations. S/N refers to 6500 \AA.}
\resizebox{0.48\textwidth}{!}{\begin{tabular}{cccccc}
\toprule
Date & HJD & System & Mid-Phase & Exposure & S/N \\
 & +2459000 &  &  & (s) & \\
\hline
2021/02/24 & 270.36145 & V455 Aur & 0.903 & 1800 & 107\\
2021/03/05 & 279.23978 & RS Ari & 0.733 & 7200 & 62\\
2021/03/29 & 303.27457 & AN Cam & 0.129 & 7200 & 68\\
\bottomrule
\label{tab:SpecObs}
\end{tabular}}
\end{table}

\section{FUNDAMENTAL PARAMETERS}

\subsection{Spectroscopic Orbit and LC Modelling}

Even though spectroscopic orbits of the three systems have already been resolved using RV data, simultaneous solutions of both RV and photometric data were carried out by using the Wilson-Devinney code \citep[WD,][]{1971ApJ...166..605W,1979ApJ...234.1054W,1990ApJ...356..613W,2008ApJ...672..575W,2010ApJ...723.1469W,2014ApJ...780..151W} to obtain fundamental parameters such as mass, radius, temperature ratio and light contributions of components for all three systems. Since these systems are detached, mode 2 of the code has been used. During the analysis, we fixed the temperature of the primary component, which is selected as the one that has higher temperature. Orbital parameters that have been obtained via spectroscopic orbits such as orbital period ($P$), mass ratio ($q$), eccentricity ($e$), and longitude of periastron ($w$) are also fixed during analysis. The adjusted parameters are conjunction time $T_0$, semi-major axis ($a$), systemic velocity ($V_\gamma$), the orbital inclination ($^{\circ}$), the temperature of secondary component ($T_2$), dimensionless surface potentials of both components ($\Omega_{1,2}$), and monochromatic luminosity ($L_1$).

\begin{figure}
    \centering
    \includegraphics[width=0.45\textwidth]{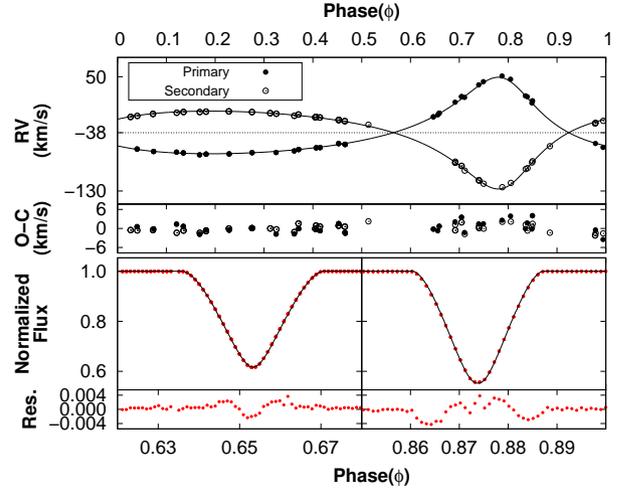}
    \caption{Models for solution of AN Cam}
    \label{fig:rvlcancam}
\end{figure}

\begin{figure}
    \centering
    \includegraphics[width=0.45\textwidth]{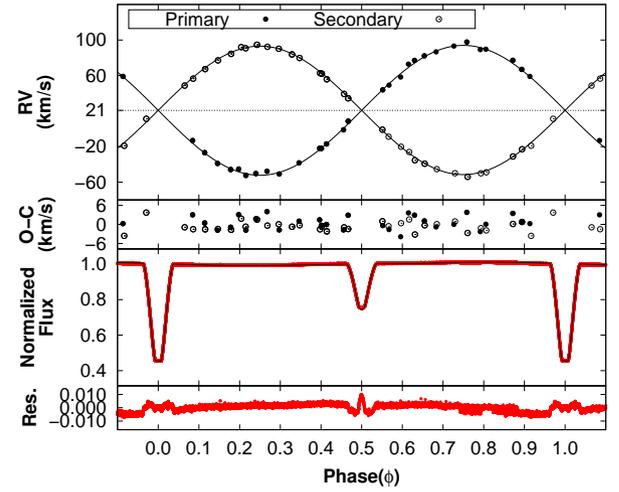}
    \caption{Models for solution of RS Ari}
    \label{fig:rvlcrsari}
\end{figure}

\begin{figure}
    \centering
    \includegraphics[width=0.45\textwidth]{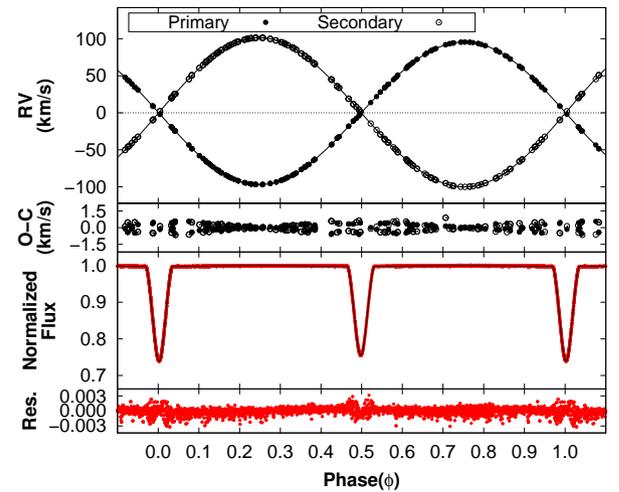}
    \caption{Models for solution of V455 Aur}
    \label{fig:rvlcv455aur}
\end{figure}

It should be noted that for V455 Aur, we have adjusted the light contribution of the third body and found it as $l_3$=0.032$\pm$0.004, which is a close value to 0.028$\pm$0.002 given by \citetalias{Southworth2021b}. Photometric and spectroscopic orbit models for the three systems are given in Figs.\ref{fig:rvlcancam}, \ref{fig:rvlcrsari}, and \ref{fig:rvlcv455aur} for AN Cam, RS Ari, and V455 Aur, respectively. In Table \ref{tab:funpar}, we present the parameters obtained as a result of the analysis. Since stellar spots are detected on the components of RS Ari, the spot features are also given in Table \ref{tab:funpar}.

The systematics in the middle of eclipses for all systems are noteworthy. For AN~Cam, the TESS magnitude is $m_{TESS}=$9.01 mag \citep[][]{Stassun2019} and the RMS of the systematics is 2 millimag (mmag) which is about 10 times bigger than the precision of the TESS data ($\sigma\sim200$ ppm) for such a brightness \citep[][]{Oelkers2018}. Nevertheless, the out of eclipse data show O--C RMS of 0.5 mmag corresponding to 500 ppm. Noticing the same systematic in AN Cam LC solution, \citetalias{Southworth2021} stated that it may be due to inhomogeneities on the components' surfaces such as star spots. In order to see the magnitude of the effect of systematic on the fitting parameters of AN Cam, we added artificial noise to the TESS data with a similar amplitude of the residuals and re-analyzed the TESS data. The resulting stellar radius parameter changed as follows: $\delta R_1=$ 0.4 percent and $\delta R_2=$ 0.3 percent. For RS~Ari, the TESS magnitude is $m_{TESS}=$9.305 mag \citep[][]{Stassun2019} and the RMS of the systematics is 3 mmag which is about 15 times bigger than the precision of the TESS data ($\sigma\sim200$ ppm) for such a brightness \citep[][]{Oelkers2018}. A similar approach is applied to see the effect of systematics on the model parameters, especially the radius of components. The artificially added noise on the TESS data yielded a change in the stellar radius as follows: $\delta R_1=$ 0.3 percent and $\delta R_2=$ 0.6 percent. For V455~Aur, the TESS magnitude is $m_{TESS}=$6.814 mag \citep[][]{Stassun2019} and the RMS of the systematics in the eclipse region is 1 mmag which is close to RMS of out of eclipse residuals (0.6 mmag) and about 13 times bigger than the precision of the TESS data ($\sigma\sim80$ ppm) for such a brightness \citep[][]{Oelkers2018}. Applying similar approach yielded a change in the stellar radius as follows: $\delta R_1=$ 0.07 percent and $\delta R_2=$ 0.4 percent. We see that the changes in radius of components for three systems are within the uncertainty limits of parameters listed in Table~\ref{tab:funpar}.

The logarithmic limb-darkening (LD) law was adopted during the LC solutions. In order to see if the selection of LD law has an effect on the improvement of LC solutions, especially to reduce systematics, linear and square root laws have also been tested. Since the LD coefficients for TESS passband are not available in  \cite{vanhamme1993}, we adopted the closest passband of Cousins\,I. It has been seen that the selection of LD law has no significant contribution to reduce the systematics seen near the eclipses of program stars. We suspect that using LD coefficients of Cousins-I band instead of TESS passband may cause low order systematics.

\subsection{Determination of Temperature and Metallicity}

Although modern photometric calibrations for temperature and metallicity determination are introduced, the spectral analysis still gives direct and the most reliable results. Hence, the temperature of each component and metallicity of each system were acquired from the observed spectra fitted by synthetic spectra which are calculated using {\sc atlas9} \citep[][]{Kurucz1979, Castelli}, and {\sc spectrum} \citep[][]{Gray} codes. By using the component light contributions, the synthetic spectrum of each component has been combined to have a composite spectrum of the system. H$_{\beta}$  region was selected for temperature determination. We had built several synthetic spectra with an interval of 100 K for AN Cam and RS Ari, and an interval of 50 K for V455 Aur since its S/N is better than others, which provides smaller uncertainties.

After the temperatures of components for each system have been obtained, several synthetic spectra with different metallicities, in an interval of 0.1 dex, with calculated temperatures have been produced. We have selected 5 regions for metallicity determination, including H$_{\beta}$ region. For each spectral region, metallicity values corresponding to minimum  ${\chi}^{2}$ were determined. Then, we have calculated the weighted average of the metallicity ($\Bar{Z}_w$) in Eq.~\eqref{average} by adopting weights as $1/{\chi}^{2}$ for each model. In Eq.~\eqref{average}, $Z$ refers to the metallicity value for each region, while $w$ and $n$ refer to the weight and number of regions, respectively.

\begin{equation}
    \Bar{Z}_w=\frac{\displaystyle\sum_{i=1} ^{n} Z_i w_i}{\displaystyle\sum_{i=1} ^{n} w_i}
    \label{average}
\end{equation}

\begin{figure*}
    \centering
    \centering
    \begin{subfigure}
        \centering
        \includegraphics[width=0.32\textwidth]{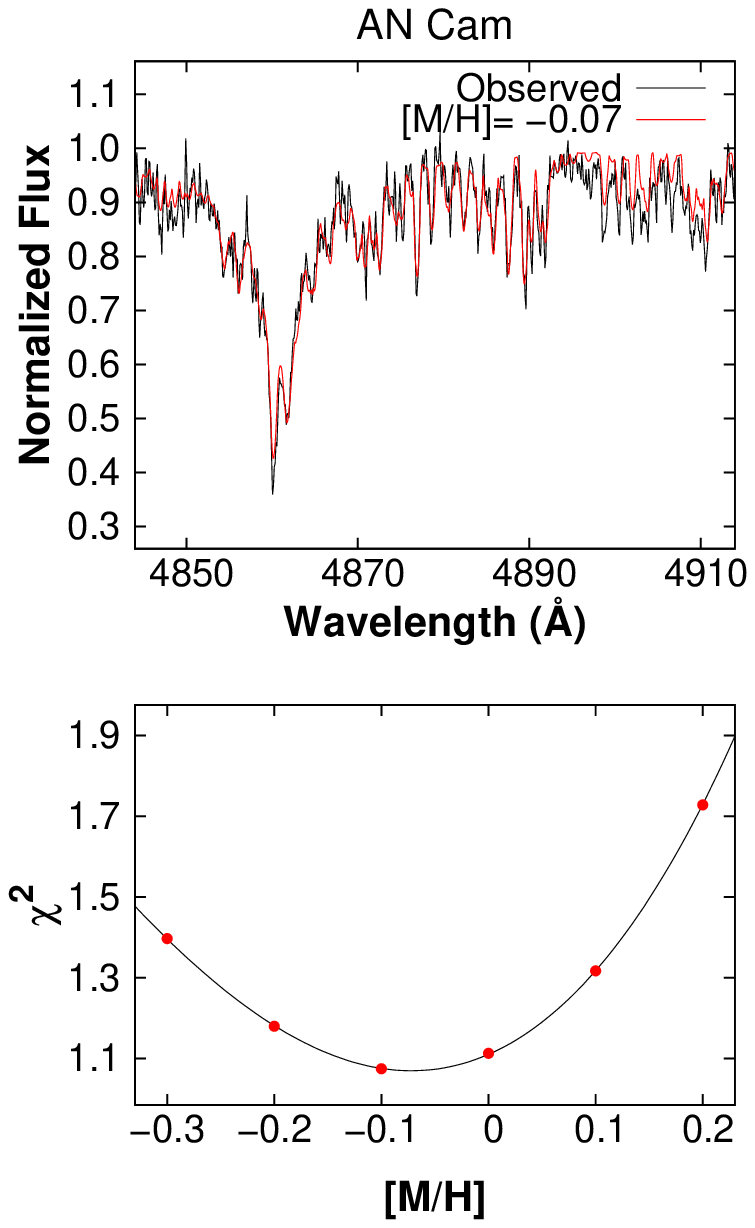}
    \end{subfigure}
    \begin{subfigure}
        \centering
        \includegraphics[width=0.32\textwidth]{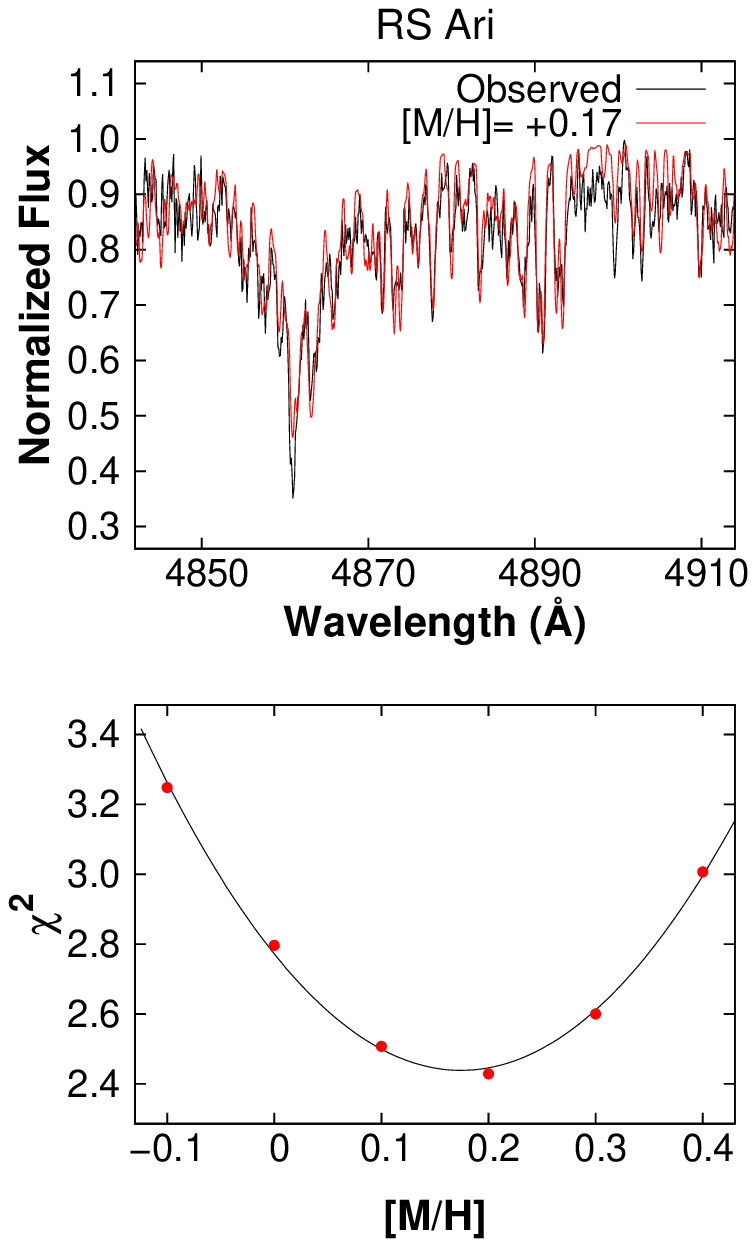}
    \end{subfigure}
    \begin{subfigure}
        \centering
        \includegraphics[width=0.32\textwidth]{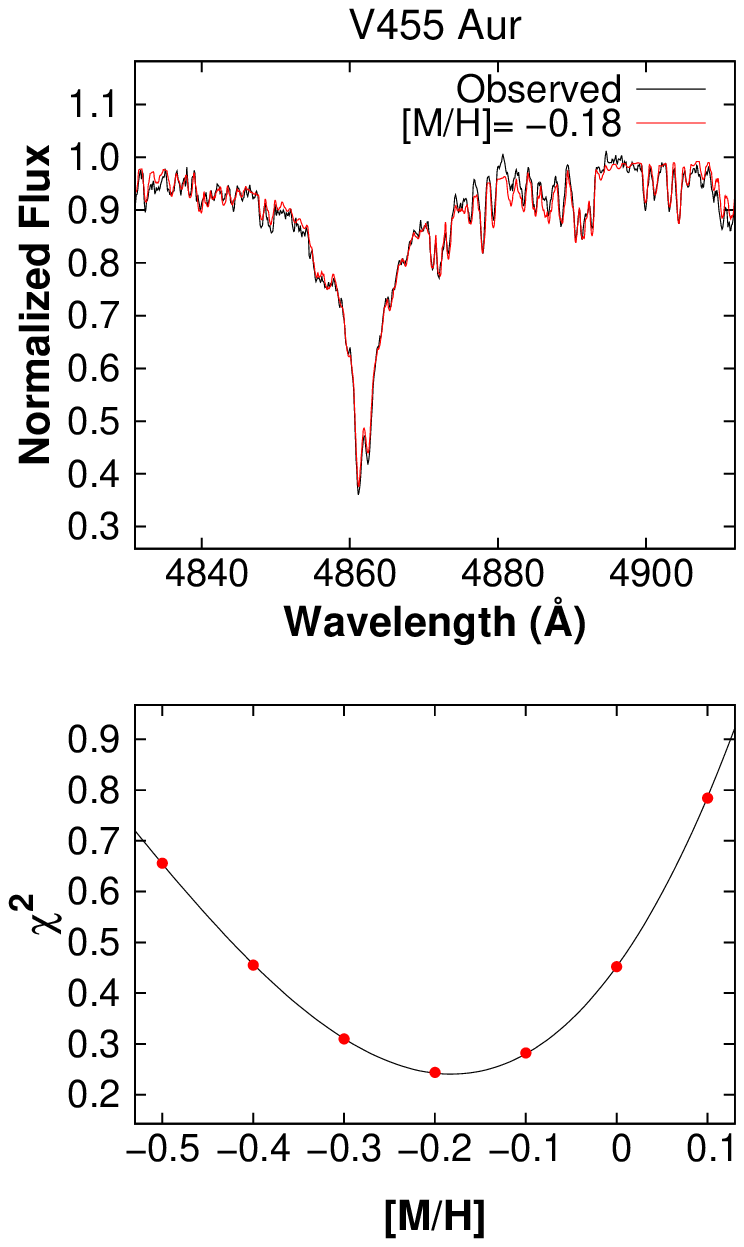}
    \end{subfigure}
    \caption{H$_\beta$ regions of three system with best model. From left to right AN Cam, RS Ari, and V455 Aur}
    \label{fig:hbeta}
\end{figure*}

 The weighted standard deviation was calculated using Eq.~\eqref{deviation} where $\sigma_w$ is the weighted standard deviation, $Z_i$ is the determined metallicity for each region, $\Bar{Z}_w$ is the determined weighted metallicity, $w$ is the weight for each region, and $n$ is the number of regions. As an example, H$_\beta$ regions of the three systems are presented in Fig.\ref{fig:hbeta}.

\begin{equation}
    \sigma_w=\sqrt{\frac{\displaystyle\sum_{i=1} ^{n} w_i \left(Z_i-\Bar{Z}_w\right)^2}{\frac{\left(n-1\right)\displaystyle\sum_{i=1} ^{n} w_i}{n}}}
    \label{deviation}
\end{equation}

\subsection{Astrophysical parameters and the distances of three systems}

As explained in the previous section, combining precise RV data with accurate photometric data made us reach the fundamental parameters of the analyzed systems with a precision better than $\sim$2 percent. 

Calculating the temperatures of components of each system enables us to determine the spectral types and intrinsic colours of the components. By using Solar absolute visual magnitude (4.75 mag) and bolometric corrections from \cite{2020eker}, bolometric magnitudes and absolute magnitudes of the components of each system have been derived. Thus, photometric distances of each system have been carried out using the distance modulus.

The calculated photometric distances of three systems are in excellent agreement with \textit{Gaia} parallax (see Table \ref{tab:funpar}), which implies that our solutions are on the mark. 

As the photometric data of three systems are from the TESS mission, LC model parameters such as relative radii of the components have been obtained with better than 0.1 per cent. However, this is not the case when the solutions have been made with RV data. In the case of simultaneous solutions the uncertainty of parameters are on the order of 1-3 per cent. Therefore, our error estimations of parameters (except temperature) are dominated by the uncertainty of spectroscopic orbital parameters, mainly semi-major axis. The code {\sc wd} estimates the errors based on dependence of $\chi^2$ on the model parameters in the region of the model fit. This is an underestimated approach in terms of calculating errors. However, according to Cramer-Rao theorem \citep[][]{aitken1942xv,frechet1943,albrecht2009findings}, any unbiased estimator for the parameters will deliver a covariance matrix on the parameters that is no better than this \citep[][]{Maxted2020}. Therefore, the error propagation is done as the following; minimum and maximum values of each parameter are calculated by considering the minimum and maximum values of its dependent parameters. Thus, the plus and minus uncertainties of the respective parameter are the differences of the parameter from the maximum and minimum values, respectively. Only, surface temperatures of the components and their uncertainties are determined directly from the spectroscopy by using grid of synthetic spectroscopic models.

\begin{table*}
\centering
\caption{Results of analysis, and fundamental parameters of AN Cam, RS Ari, and V455 Aur, respectively. }
\resizebox{\textwidth}{!}{\begin{tabular}{lcccccc}
\toprule
\multirow{2}*{Parameter} & \multicolumn{2}{c}{AN~Cam} & \multicolumn{2}{c}{RS~Ari} & \multicolumn{2}{c}{V455~Aur}\\
 & Primary & Secondary & Primary & Secondary & Primary & Secondary \\
\hline
\textbf{\textsc{wilson-devinney} Analysis} \\
	$P$(days)      &   \multicolumn{2}{c}{20.99846} & \multicolumn{2}{c}{8.803172} & \multicolumn{2}{c}{3.145799} \\
	$T_{0}$(HJD-2400000)   & \multicolumn{2}{c}{58826.6769 $\pm$ 0.0003} & \multicolumn{2}{c}{58802.2524  $\pm$ 0.0008} & \multicolumn{2}{c}{58858.5671 $\pm$ 0.0002}\\
	$K_{1,2}$(km s$^{-1}$) & 62.03 $\pm$ 0.31 &  61.01 $\pm$ 0.28 & 73.10 $\pm$ 0.47 & 72.28 $\pm$ 0.43 & 96.2066 $\pm$ 0.06 & 100.6080 $\pm$ 0.06\\
	$e$        & \multicolumn{2}{c}{0.47001 $\pm$ 0.00028} & \multicolumn{2}{c}{0 (fixed)} & \multicolumn{2}{c}{0.00930 $\pm$ 0.00029} \\
	$w$ ($^o$)        & \multicolumn{2}{c}{194.4 $\pm$ 0.5} & \multicolumn{2}{c}{0 (fixed)} & \multicolumn{2}{c}{132.5 $\pm$ 1.9} \\
	$V_{\gamma}$(km s$^{-1}$)   &  \multicolumn{2}{c}{-38.00 $\pm$ 0.14} & \multicolumn{2}{c}{20.81 $\pm$ 0.24} & \multicolumn{2}{c}{0 (fixed)}\\
	$q (M_2/M_1)$    &   \multicolumn{2}{c}{1.02 $\pm$ 0.01} & \multicolumn{2}{c}{1.01 $\pm$ 0.01} & \multicolumn{2}{c}{0.96 $\pm$ 0.01} \\
	$M_{1,2}{\mathrm sin}^{3}i (M_{\odot})$ &  1.382 $\pm$ 0.023 &  1.405 $\pm$ 0.022 & 1.397 $\pm$ 0.026 & 1.412 $\pm$ 0.027 & 1.270 $\pm$ 0.002 &  1.214 $\pm$ 0.002 \\
	$a {\mathrm sin}i (R_{\odot})$     & \multicolumn{2}{c}{45.058 $\pm$ 0.215} & \multicolumn{2}{c}{25.286 $\pm$ 0.161} & \multicolumn{2}{c}{12.162 $\pm$ 0.004}\\
	$T_{\mathrm{eff}1}/T_{\mathrm{eff}2}$  & \multicolumn{2}{c}{1.025 $\pm$ 0.003} & \multicolumn{2}{c}{0.833 $\pm$ 0.009} & \multicolumn{2}{c}{1.012 $\pm$ 0.001} \\
	$i$ ($^o$) & \multicolumn{2}{c}{88.972 $\pm$ 0.006} & \multicolumn{2}{c}{88.725 $\pm$ 0.031} & \multicolumn{2}{c}{84.971 $\pm$ 0.001} \\
	$l_1/l_{\mathrm{total}}$ & \multicolumn{2}{c}{0.449 $\pm$ 0.005} & \multicolumn{2}{c}{0.461 $\pm$ 0.002} & \multicolumn{2}{c}{0.508 $\pm$ 0.001} \\
	$l_3/l_{\mathrm{total}}$ & & & & & \multicolumn{2}{c}{0.032 $\pm$ 0.004} \\
	Limb-darkening $X_{1,2}$ & 0.543 & 0.563 & 0.556 & 0.633 & 0.532 & 0.539\\
	Limb-darkening $Y_{1,2}$ & 0.267 & 0.258 & 0.262 & 0.169 & 0.268 & 0.266\\
	$\Omega_{1,2}$ & 22.370 $\pm$ 0.007 & 19.041 $\pm$ 0.009 & 11.246 $\pm$ 0.015 & 8.125 $\pm$ 0.006 & 9.711 $\pm$ 0.001 & 9.839 $\pm$ 0.001\\
	$r_{1,2}$ & 0.049 $\pm$ 0.001 & 0.059 $\pm$ 0.001 & 0.095 $\pm$ 0.006 & 0.139 $\pm$ 0.003 & 0.115 $\pm$ 0.001 & 0.109 $\pm$ 0.001\\
	\textbf{Spot Parameters} \\
	Co-latitude ($^o$) & & & 90 $\pm$ 1 & 88 $\pm$ 1 \\
	Longitude ($^o$) & & & 270 $\pm$ 1 & 260 $\pm$ 1 \\
	Radius ($^o$) & & & 10 $\pm$ 1 & 19 $\pm$ 1 \\
	Temperature Ratio & & & 0.74 $\pm$ 0.05 &  1.02 $\pm$ 0.02 \\ 
	\textbf{Spectral Analysis} \\
	$T_{\mathrm{eff}1,2}$ (K) & 6050 $\pm$ 100 & 5900 $\pm$ 100 & 6000 $\pm$ 100 & 5000 $\pm$ 100 & 6500 $\pm$ 50 & 6424 $\pm$ 50 \\
	\text{[M/H]} & \multicolumn{2}{c}{0.00 $\pm$ 0.12} &\multicolumn{2}{c}{0.05 $\pm$ 0.08} & \multicolumn{2}{c}{-0.07 $\pm$ 0.07} \\
	$v {\mathrm {sin}} i_{1,2}$ (km s$^{-1}$) & 25 $\pm$ 5 & 25 $\pm$ 5 & 20 $\pm$ 5 & 20 $\pm$ 5 & 25 $\pm$ 5 & 25 $\pm$ 5\\
	\textbf{Fundamental Parameters} \\
	Spectral Type (Sp) & F9.5 IV & G1.5 IV & G0 IV & K1.5 III & F4 V & F4.5 V \\  
	M$_{1,2}$ ($\mathrm{M}_\odot$) & 1.383 $\pm$ 0.025 & 1.406 $\pm$ 0.024 & 1.398 $\pm$ 0.028 & 1.413 $\pm$ 0.029 & 1.287 $\pm$ 0.003 & 1.231 $\pm$ 0.003\\
	R$_{1,2}$ ($\mathrm{R}_\odot$) & 2.206 $\pm$ 0.059 & 2.667 $\pm$ 0.058 & 2.403 $\pm$ 0.041 & 3.531 $\pm$ 0.048 & 1.409 $\pm$ 0.013 & 1.339 $\pm$ 0.013\\
	a ($\mathrm{R}_\odot$) & \multicolumn{2}{c}{45.065 $\pm$ 0.252} & \multicolumn{2}{c}{25.312$\pm$0.160} & \multicolumn{2}{c}{12.286 $\pm$ 0.007} \\
	log $g_{1,2}$ (cm s$^{-2}$) & 3.892 $\pm$ 0.059 & 3.734 $\pm$ 0.055 & 3.822 $\pm$ 0.044 & 3.492 $\pm$ 0.040 & 4.250 $\pm$ 0.009 & 4.275 $\pm$ 0.010\\
	log L$_{1,2}$/L$_\odot$ & 0.787 $\pm$ 0.036 & 0.891 $\pm$ 0.034 & 0.829 $\pm$ 0.044 & 0.847 $\pm$ 0.047 & 0.505 $\pm$ 0.021 & 0.440 $\pm$ 0.022\\
	Combined visual magnitude (mag)\footnote{taken from SIMBAD} & \multicolumn{2}{c}{9.69 $\pm$ 0.02} &\multicolumn{2}{c}{10.02 $\pm$ 0.04} & \multicolumn{2}{c}{7.28 $\pm$ 0.01} \\
	Individual visual magnitudes (mag) & 10.360 $\pm$ 0.011 & 10.137 $\pm$ 0.011 & 10.391 $\pm$ 0.040 & 10.561 $\pm$ 0.047 & 7.894 $\pm$ 0.010 & 8.191 $\pm$ 0.013\\
	Combined colour index $(B-V)$(mag)\footnote{taken from SIMBAD} & \multicolumn{2}{c}{0.61 $\pm$ 0.05} & \multicolumn{2}{c}{0.82 $\pm$ 0.10} & \multicolumn{2}{c}{0.43 $\pm$ 0.02} \\
	Bolometric magnitude (mag) & 2.773 $\pm$ 0.078 & 2.513 $\pm$ 0.083 & 2.678 $\pm$ 0.110 & 2.633 $\pm$ 0.118 & 3.488 $\pm$ 0.053 & 3.650 $\pm$ 0.055\\
	Absolute visual magnitude (mag) & 2.766 $\pm$ 0.071 & 2.483 $\pm$ 0.083 & 2.655 $\pm$ 0.110 & 2.890 $\pm$ 0.075 & 3.417 $\pm$ 0.050 & 3.584 $\pm$ 0.055\\
	Bolometric correction (mag) & 0.029 $\pm$ 0.007 & 0.007 $\pm$ 0.006 & 0.023 $\pm$ 0.016 & -0.257 $\pm$ 0.043 & 0.071 $\pm$ 0.003 & 0.066 $\pm$ 0.003\\
	Computed synchronization velocities (km s$^{-1}$) & 6.2 $\pm$ 0.1 & 5.6 $\pm$ 0.1 & 13.8 $\pm$ 0.2 & 20.3 $\pm$ 0.3 & 22.7 $\pm$ 0.2 & 21.5 $\pm$ 0.2\\
	Distance (pc) & \multicolumn{2}{c}{326 $\pm$ 12} & \multicolumn{2}{c}{338 $\pm$ 19} & \multicolumn{2}{c}{75 $\pm$ 4}\\
	\textit{Gaia} Distance (pc) & \multicolumn{2}{c}{311 $\pm$ 1} & \multicolumn{2}{c}{337 $\pm$ 2} & \multicolumn{2}{c}{77 $\pm$ 1} \\
\bottomrule
$T_{\mathrm{0}}$ refers to time of periastron passage.\\
\footnotesize{${^4}{^,}{^5}$ Taken from \citet[]{Hog}}\\
\label{tab:funpar}
\end{tabular}}
\end{table*}

\section{Evolutionary Analysis}

One of the main goals of this paper is to test our results with up-to-date theoretical evolutionary tracks. In that regard, we have used version 12115 of Modules for Experiments in Stellar Astrophysics
\citep[MESA,][]{Paxton2011, Paxton2013, Paxton2015, Paxton2018, Paxton2019} to calculate the evolution of these systems. We have done that procedure in two sections, single star evolution and binary star evolution.

\subsection{Single Star Evolution}

\begin{figure*}
    \centering
    \centering
    \begin{subfigure}
        \centering
        \includegraphics[width=0.30\textwidth]{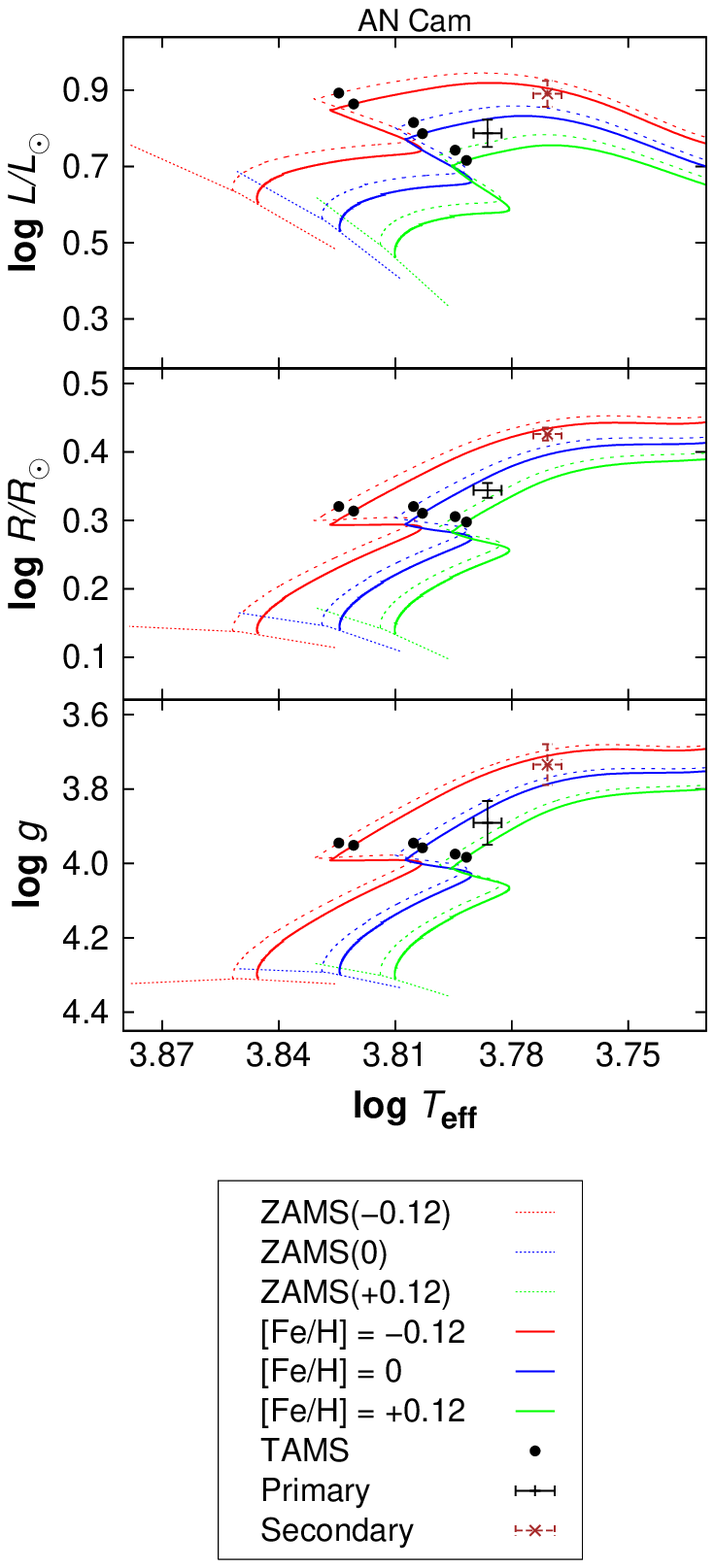}
    \end{subfigure}
    \begin{subfigure}
        \centering
        \includegraphics[width=0.30\textwidth]{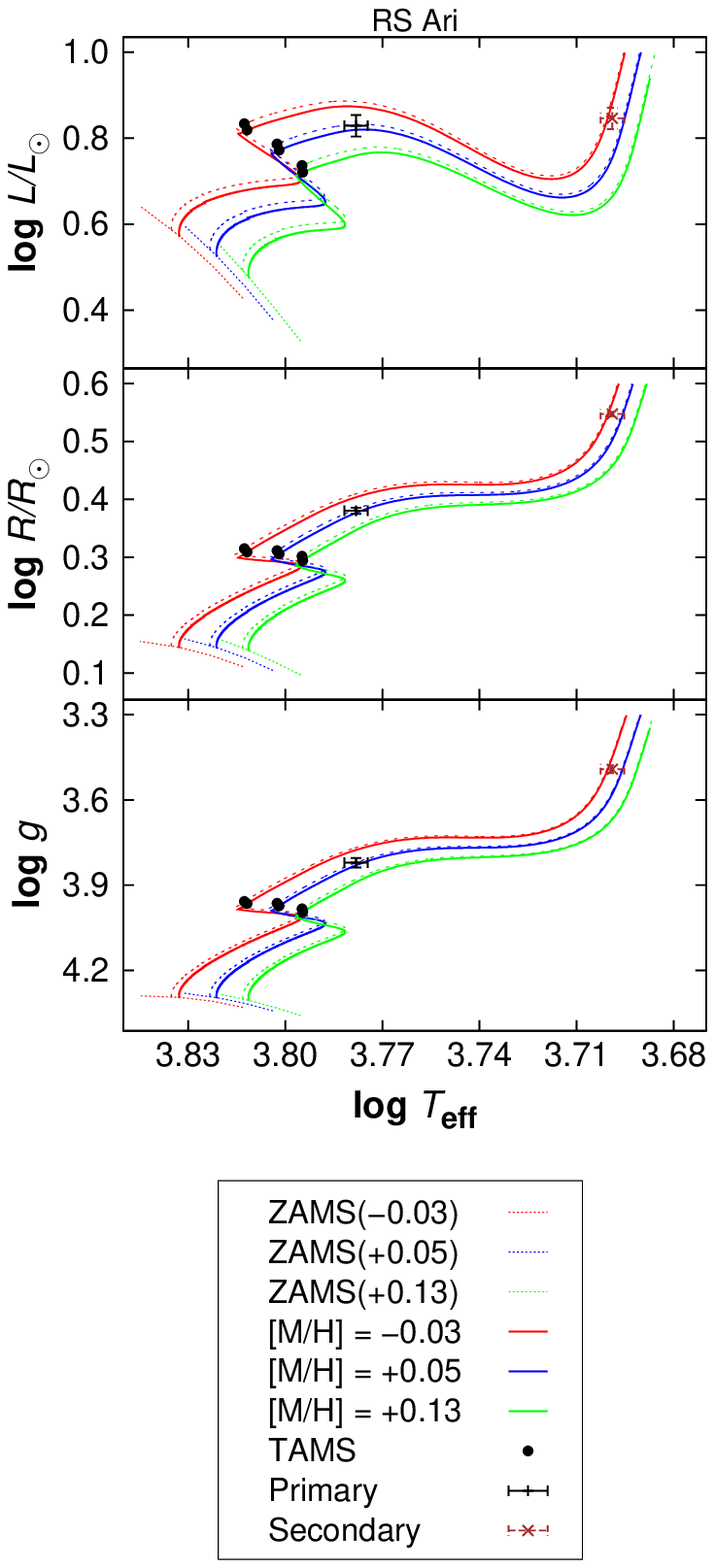}
    \end{subfigure}
    \begin{subfigure}
        \centering
        \includegraphics[width=0.30\textwidth]{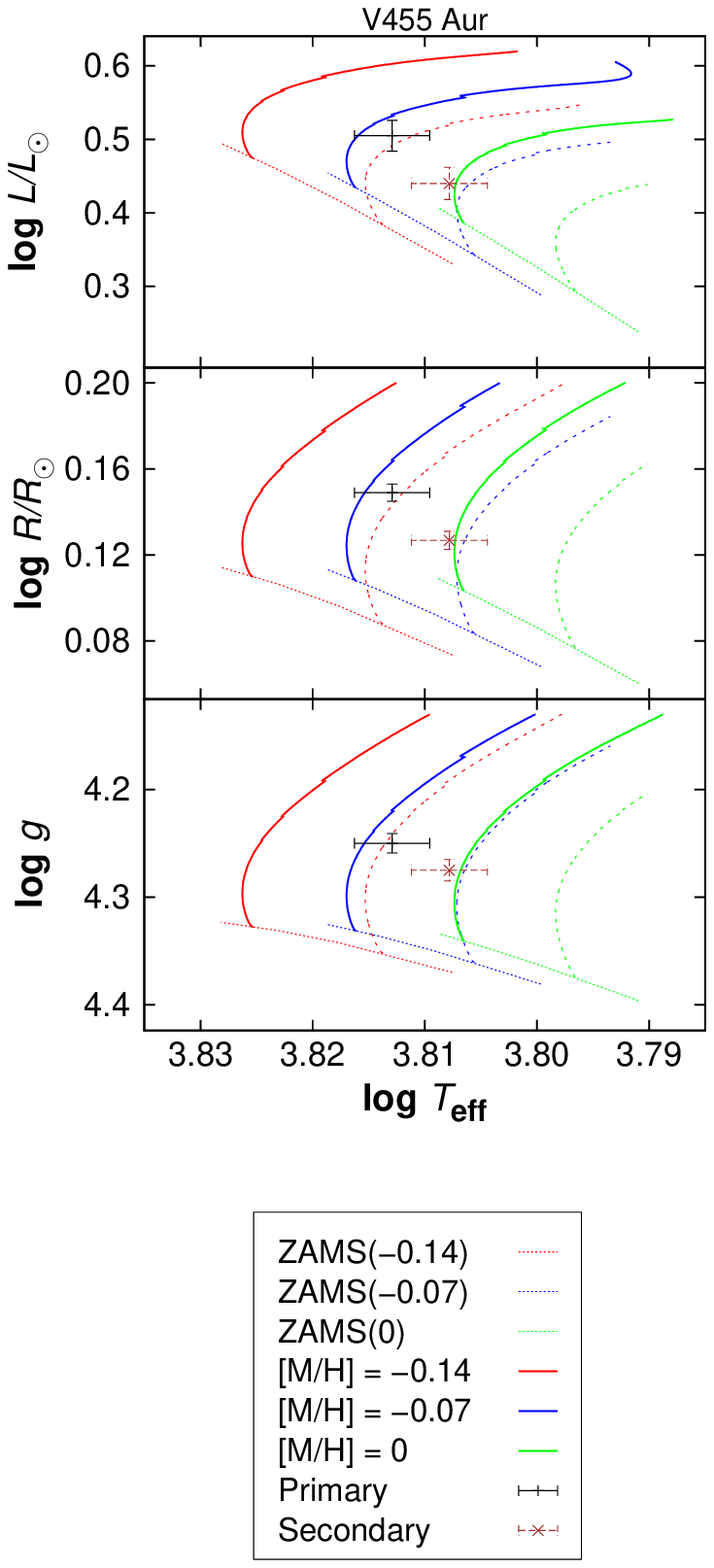}
    \end{subfigure}
    \caption{Single star evolutionary tracks for three systems. From left to right: AN Cam, RS Ari, and V455 Aur.}
    \label{fig:single}
\end{figure*}

Components of the systems we analyzed are still in their Roche lobes, which means that there is no mass transfer in the systems. Thus, components of the systems can be analyzed as a single star. We have calculated evolutionary tracks by using mass and metallicity, in the range of uncertainties, with a single rotating star as the component of each system. We have also calculated evolutionary tracks of several different masses with measured metallicities by using spectral data to build Zero Age Main Sequence (ZAMS) lines and Terminal Age Main Sequence (TAMS) lines with each metallicity value with uncertainty, which are represented by dashed lines and black dots, respectively. Our results are given in  Fig.\ref{fig:single} as log $T_{\mathrm{eff}}$-log $L$, log $T_{\mathrm{eff}}$-log $R$/$R_\odot$, and log $T_{\mathrm{eff}}$-log$g$ planes. Continuous lines represent the primary components, and dashed lines represent the secondary components in each system, respectively. Metallicities are represented by associated colour. According to our analysis, and as can be seen in Fig.\ref{fig:single}, the primary component of AN Cam is in the phase of thin hydrogen shell burning around its helium core while the secondary one is almost finishing thick hydrogen shell burning around its helium core and will start thin hydrogen shell burning around its helium core. The primary component of RS Ari is in the phase of becoming a red giant while the secondary component of RS Ari has just started to burn thin shell hydrogen around its helium core. Finally, both components of V455 Aur are on the main-sequence and still burning hydrogen in their cores.

\subsection{Binary Stars Evolution}

Since the three systems are detached, a single star evolution approximation is suitable. But in need of understanding the processes during the evolution of each system, one must calculate the evolution from the start, which requires the inclusion of initial orbital parameters. \textsc{mesa} provides this opportunity with its binary module, which enables the evolution of both components in a system with relevant orbital changes with respect to given initial period, or semi-major axis, and eccentricity values. Since these systems are detached, there has been no change in masses due to mass transfer to this day. Therefore, initial masses in the systems have the same value as they had started their evolution, with only two parameters that have changed over time due to orbital evolution, which are period and eccentricity. For finding initial period and eccentricity values for each system, a method similar to \cite{Rosales}, has been used. Several model calculations were made for each system with different initial periods and initial eccentricity values with a stopping condition of the current eccentricity value of that system. When the model has reached to current eccentricity value for each system, the evolution is stopped and a $\chi^2$ calculation has been made, which includes calculated radius and temperature of the components, and period with the current value of corresponding parameters of each system. In our model calculations, for the angular momentum change, we activated the setting of magnetic braking \citep[][]{Rappaport}, for tidal synchronization, we had used "Orb$\_$period" option, which synchronizes the orbit relevant to the timescale of the orbital period. We also applied tidal circularisation, given by \cite{Hurley}, on both components of AN Cam and V455 Aur since these systems have eccentric orbits. Binary evolution calculations were made by using calculated metallicity values. 

\subsubsection{AN Cam}

\begin{figure}
    \centering
    \includegraphics[width=0.47\textwidth]{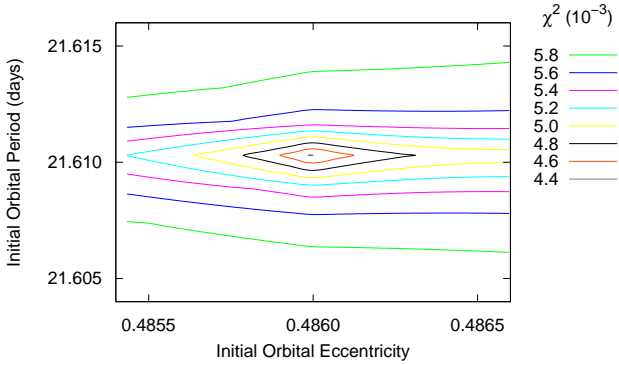}
    \caption{$\chi^2$ map of corresponding initial period for each initial orbital eccentricity of AN Cam.}
    \label{fig:ancamchi}
\end{figure}

\begin{figure}
    \centering
    \centering
    \begin{subfigure}
        \centering
        \includegraphics[width=0.47\textwidth]{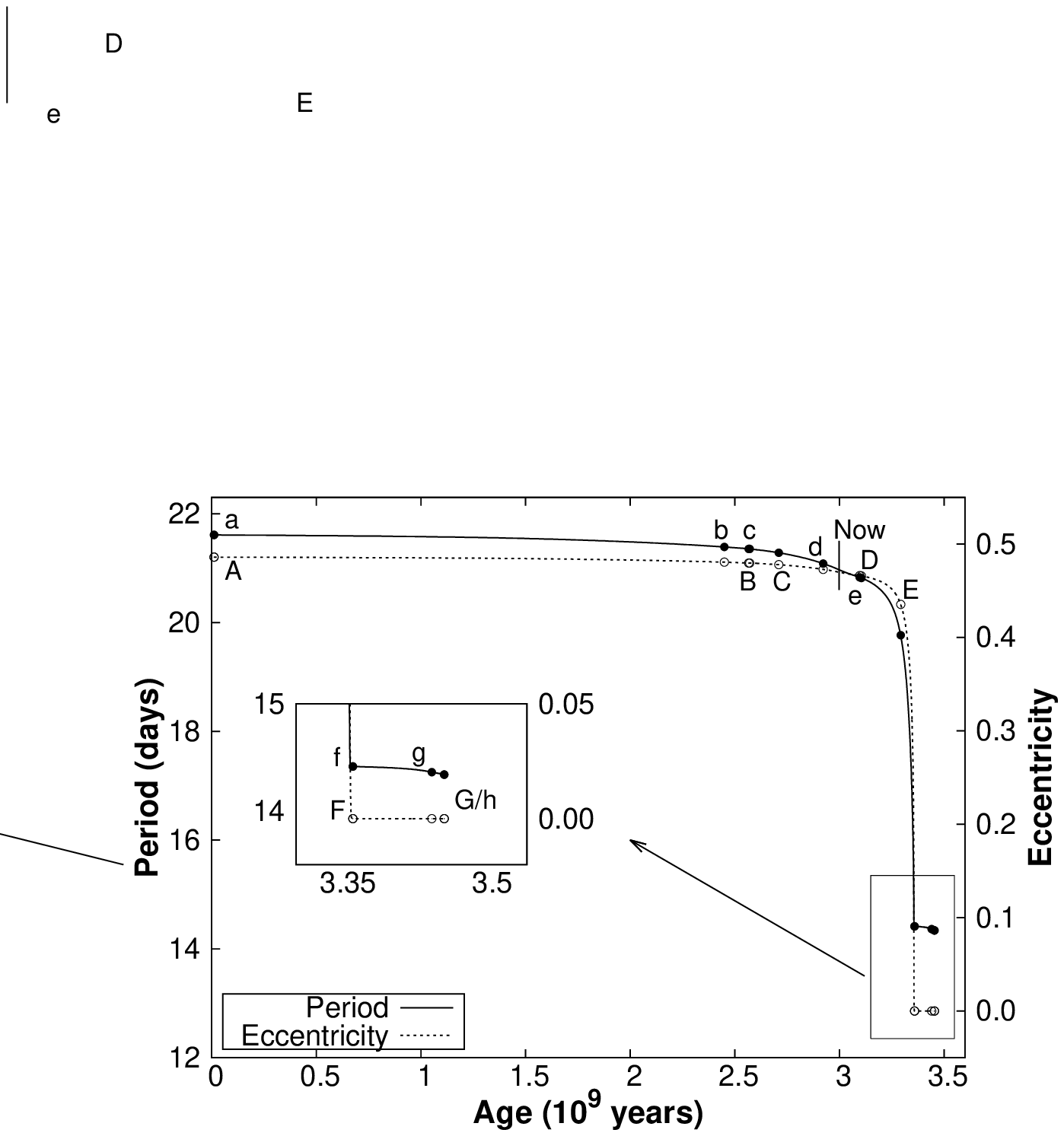}
    \end{subfigure}
    \begin{subfigure}
        \centering
        \includegraphics[width=0.47\textwidth]{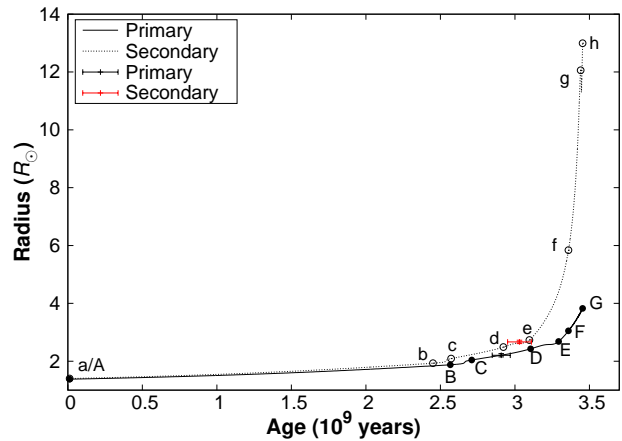}
    \end{subfigure}
    \caption{Change of orbital parameters and radius of components of AN Cam with time}
    \label{fig:mesaancam}
\end{figure}

\begin{figure}
    \centering
    \includegraphics[width=0.48\textwidth]{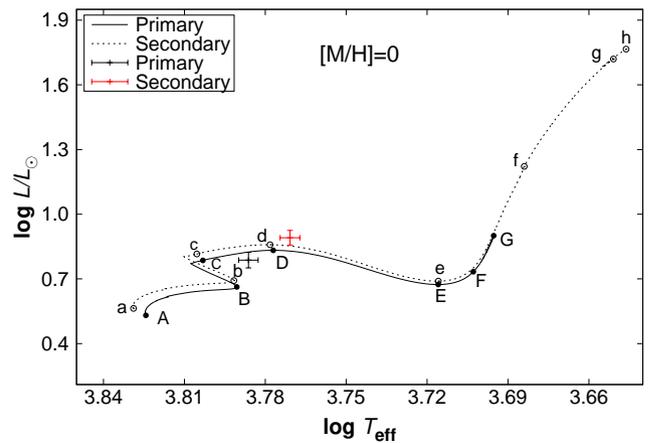}
    \caption{Evolution of AN Cam on log $T_\mathrm{eff}$-log$L/{L_\odot}$ plane}
    \label{fig:ancamhr}
\end{figure}

As explained in the previous section, we built evolution models that initial parameters change for the period and eccentricity between 21.50 and 21.80 days with an interval of 0.005 days and between 0.4820 and 0.4900 with an interval of 0.0002, respectively, and kept the evolution continued until eccentricity drops to 0.47001, which is the up-to-date eccentricity value of AN Cam. Finally, we selected the best model, which gives the smallest $\chi^2$ value of 0.0044, with initial orbital parameters for period and eccentricity as, 21.61 and 0.4860, respectively (given in Fig. \ref{fig:ancamchi}). After that, we made an evolution model with those initial orbital parameters and stopped the model until the mass transfer started through the Roche lobe. According to our model, the age of AN Cam is 3.00$\pm$0.15 Gyrs, and the primary component of AN Cam will start transferring mass through its Roche lobe when the primary component is in the phase of thermal pulses while the secondary component is in the phase of a red giant when the age of the system is 3.511 Gyrs. Changes in orbital parameters and radii of the components during the evolution are presented in Fig. \ref{fig:mesaancam}. Detailed evolution of both components of AN Cam with timetables are given in Table \ref{tab:ancamtime} and shown in Fig.\ref{fig:ancamhr}.

It is also interesting to note what happens after the mass transfer begins in the system. Considering the fact that the more massive component of AN Cam starts mass transferring to the less massive component just after the thermal pulses begins (AGB-stage) (see Table \ref{tab:ancamtime}), there will also be a non-conservative mass loss from the donor on rates changing between $10^{-8}-10^{-5} M_\odot year^{-1}$ \citep[][]{Hofner2018}. This means that the donor will have lost about 50-70 percent of its mass during this stage not only because of the mass transfer but also due to mass loss causing it to be a white dwarf (WD) in the end of its evolution. The mass transfer and loss rates play an important role in the evolution of the system. It is, therefore, difficult to predict the fate of the system without exact knowledge of these parameters. However, we adopted average rates (i.e. \cite{Paxton2015,Rosales,Soydugan}) while calculating the binary star evolution to have an idea about the end of the system. These average rates refer to $\alpha=$0.4, $\beta=$0.1 and $\gamma=$0.1 for fraction of lost from the vicinity of the donor as fast wind, fraction of lost from the vicinity of the accretor as fast wind, and fraction of lost from the circumstellar disk, respectively. These assumptions lead to a WD-giant binary with stellar masses of $M_{\mathrm{WD}}$=0.32 $M_\odot$ and $M_{\mathrm{Giant}}$=1.82 $M_\odot$. We need to have more evolution studies of semi-detached or contact binary systems in the literature \citep[among others;][]{Budding2005, HBakis2008,HBakis2016,HBakis2021} to have a more reliable knowledge about the mass transfer and loss rates in these systems for a better estimation about the end of the evolution of detached binary systems.

\begin{table*}\centering
    \caption{Detailed evolution of AN Cam with time stamps}
    \resizebox{\textwidth}{!}{\begin{tabular}{lccccccccccc}
    \toprule
        & \multirow{2}*{Mark} & \multirow{2}*{Evolutionary Status} & Age & Period & \multirow{2}*{Eccentricity} & \multicolumn{3}{c}{Primary} & \multicolumn{3}{c}{Secondary} \\
        & & & (Myears) & (days) &  & log $T_\mathrm{eff}$ (K) & log $L$ (L$_\odot$)  & Radius (R$_\odot$) & log $T_\mathrm{eff}$ (K) & log $L$ (L$_\odot$)  & Radius (R$_\odot$)\\
        \hline
        \parbox[t]{2mm}{\multirow{7}{*}{\rotatebox[origin=c]{90}{Primary}}} & A & Zero age main sequence & 0 & 21.610 & 0.4860 & 3.824 & 0.531 & 1.380 & 3.829 & 0.563 & 1.403  \\
        & B & Core contraction & 2566 & 21.354 & 0.4798 & 3.791 & 0.663 & 1.877 & 3.806 & 0.814 & 2.085 \\
        & C & Terminal age main sequence & 2710 & 21.282 & 0.4780 & 3.803 & 0.786 & 2.041 & 3.797 & 0.823 & 2.188 \\
        & D & Thin H shell burning & 3105 & 20.818 & 0.4659 & 3.777 & 0.833 & 2.429 & 3.797 & 0.837 & 2.230 \\
        & E & Entering red giant phase & 3293 & 19.766 & 0.4355 & 3.716 & 0.674 & 2.682 & 3.692 & 1.008 & 4.400\\
        & F & Circularisation of Orbit & 3359 & 14.415 & 0 & 3.703 & 0.734 & 3.050 & 3.684 & 1.222 & 5.839  \\
        & G & Starting of mass transfer & 3454 & 14.338 & 0 & 3.695 & 0.900 & 3.826 & 3.646 & 1.766 & 13.099 \\
        \midrule
        \parbox[t]{2mm}{\multirow{8}{*}{\rotatebox[origin=c]{90}{Secondary}}} & a & Zero age main sequence & 0 & 21.610 & 0.4860 & 3.824 & 0.531 & 1.380 & 3.829 & 0.563 & 1.403  \\
        & b & Core contraction & 2450 & 21.389 & 0.4806 & 3.793 & 0.653 & 1.841 & 3.792 & 0.692 & 1.932 \\
        & c & Terminal age main sequence & 2571 & 21.352 & 0.4797 & 3.791 & 0.664 & 1.879 & 3.805 & 0.815 & 2.090 \\
        & d & Thin H shell burning & 2922 & 21.084 & 0.4729 & 3.792 & 0.814 & 2.220 & 3.778 & 0.859 & 2.489\\ 
        & e & Entering red giant phase & 3097 & 20.829 & 0.4662& 3.778 & 0.832 & 2.418 & 3.716 & 0.690 & 2.731 \\
        & f & Circularisation of Orbit & 3359 & 14.415 & 0 & 3.703 & 0.734 & 3.050 & 3.684 & 1.222 & 5.839 \\
        & g & Thermal Pulses of Secondary & 3441 & 14.362 & 0 & 3.696 & 0.874 & 3.695 & 3.651 & 1.720 & 12.060 \\
        & h & Starting of mass transfer & 3454 &  14.338 & 0 & 3.695 & 0.900 & 3.826 & 3.646 & 1.766 & 13.099\\
        \bottomrule
    \end{tabular}}
    \label{tab:ancamtime}
\end{table*}

\subsubsection{RS Ari}

\begin{figure}
    \centering
    \includegraphics[width=0.47\textwidth]{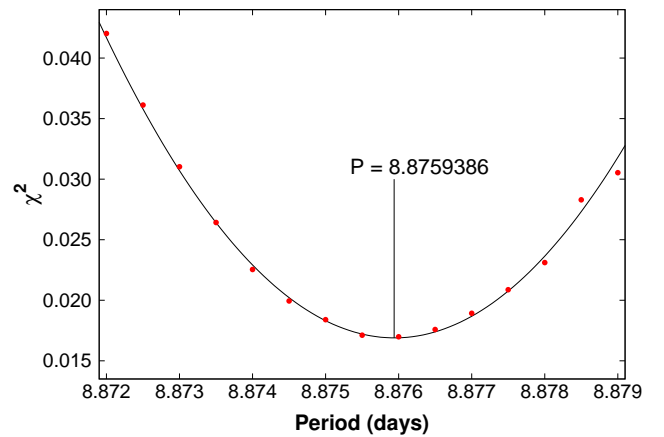}
    \caption{Fitting procedure for finding initial period of RS Ari}
    \label{fig:rsariperiod}
\end{figure}

\begin{figure}
    \centering
    \centering
    \begin{subfigure}
        \centering
        \includegraphics[width=0.47\textwidth]{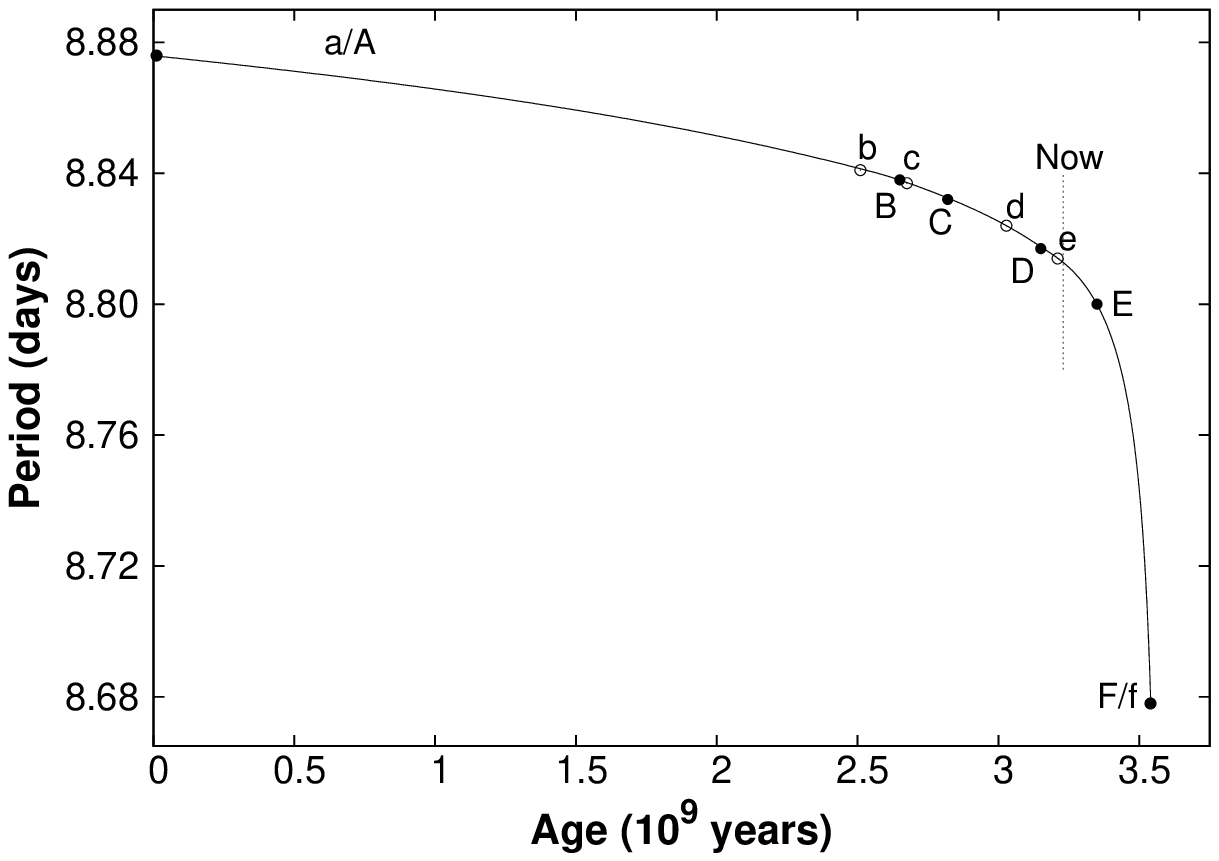}
    \end{subfigure}
    \begin{subfigure}
        \centering
        \includegraphics[width=0.47\textwidth]{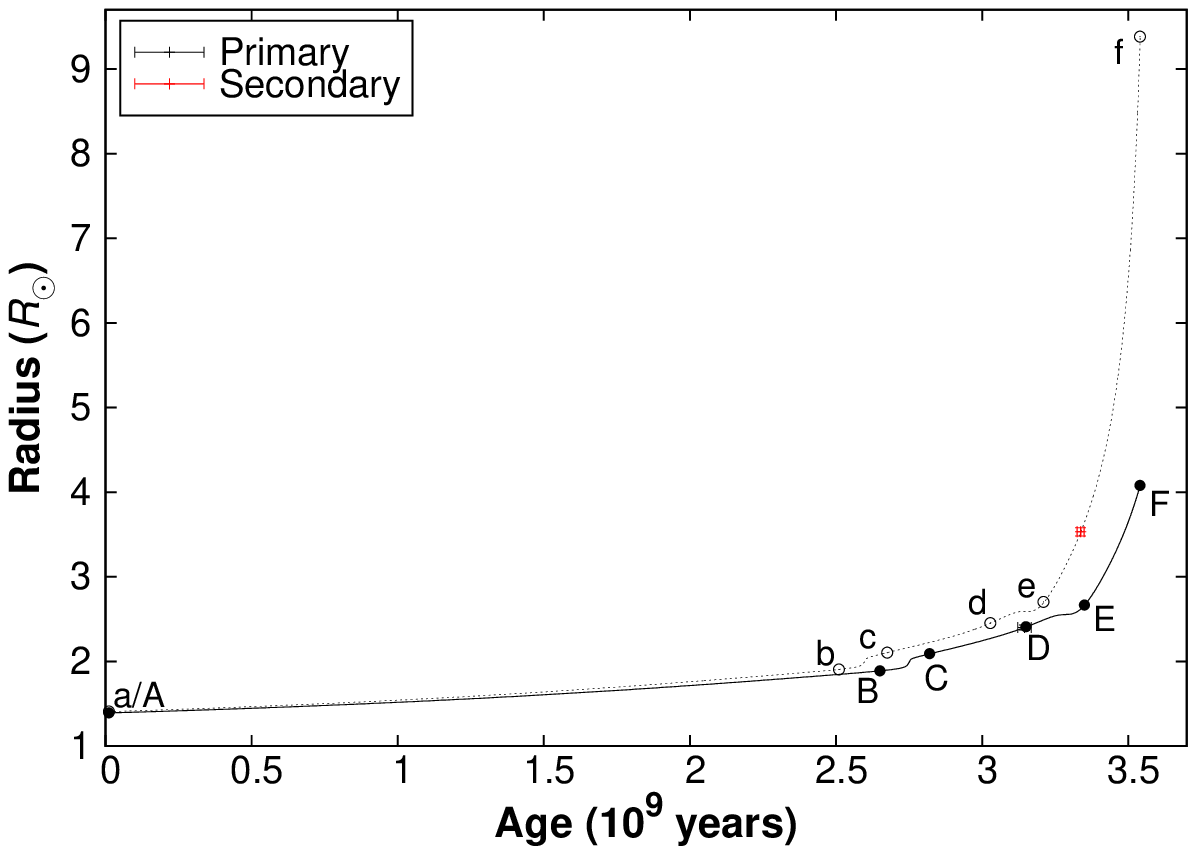}
    \end{subfigure}
    \caption{Change of period and radius of components of RS Ari in time}
    \label{fig:mesarsari}
\end{figure}

\begin{figure}
    \centering
    \includegraphics[width=0.49\textwidth]{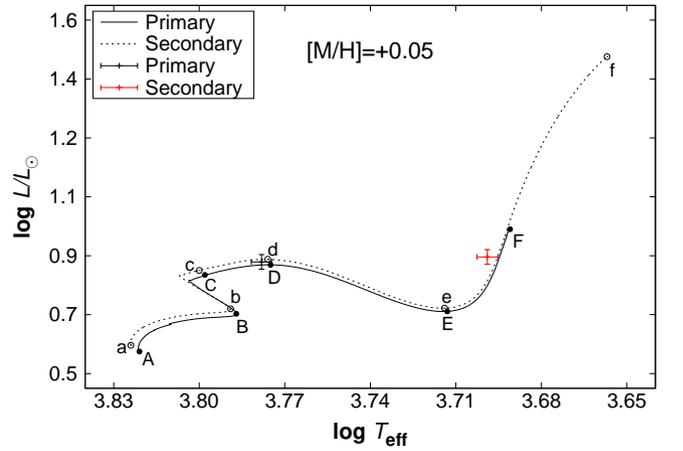}
    \caption{Evolution of RS Ari on $T_\mathrm{eff}$-log$L/{L_\odot}$ plane}
    \label{fig:rsarihr}
\end{figure}

We had used the same procedure on RS Ari, except for the fact that since there is no eccentricity in the system, there is no way to know whether the system has started its evolution with an eccentricity, if it has, when the orbit of the system has circularized, or system has started its evolution with no eccentricity. Hence, we assumed that RS Ari had started its evolution with a circular orbit; therefore, the only initial orbital parameter that would change in time is the period of the system. Therefore, we started evolution with periods between 8.87 days and 8.88 days with an interval of 0.0005 days. We have used the current period of the system as a stopping condition, which is 8.803172 days. When the models have stopped, we calculated a $\chi^2$, which was obtained by using the temperature and radius of components’ current values and values models give. Lastly, we fitted the $\chi^2$ values against the period and found the best-fitted period of the system, 8.8759386 days. In Fig.\ref{fig:rsariperiod}, fitting procedure is presented. Hence, we have used that period value as the initial period of the system and calculated the evolution model until the start of the mass transfer in the system. According to our model, the age of RS Ari is 3.23 $\pm$ 0.11 Gyrs, and the primary component of RS Ari will start transferring mass through its Roche lobe when the primary component is in the phase of a red giant, while the secondary component becomes giant when the age of the system is 3.54 Gyrs. The changes in orbital parameters and radii of components during evolution are presented in Fig.\ref{fig:mesarsari}. Detailed evolution of both components of RS Ari with timetables are given in Table \ref{tab:rsaritime} and Fig.\ref{fig:rsarihr}. Similar to AN Cam, RS Ari will become a WD-Giant binary in the late stages of its evolution. The stellar masses will be $M_{\mathrm{WD}}$=0.30 $M_\odot$ and $M_{\mathrm{Giant}}$=1.84 $M_\odot$ under the mass loss and transfer rate assumptions as taken for AN Cam.

\begin{table*}\centering
    \caption{Detailed evolution of RS Ari with time stamps}
    \resizebox{\textwidth}{!}{\begin{tabular}{lcccccccccc}
    \toprule
        & \multirow{2}*{Mark} & \multirow{2}*{Evolutionary Status} & Age & Period  & \multicolumn{3}{c}{Primary} & \multicolumn{3}{c}{Secondary} \\
        & & & (Myears) & (days) & log $T_\mathrm{eff}$ (K) & log $L$ (L$_\odot$)  & Radius (R$_\odot$) & log $T_\mathrm{eff}$ (K) & log $L$ (L$_\odot$)  & Radius (R$_\odot$)\\
        \hline
        \parbox[t]{2mm}{\multirow{6}{*}{\rotatebox[origin=c]{90}{Primary}}} & A & Zero age main sequence & 0 & 8.876  & 3.821 & 0.525 & 1.392 & 3.824 & 0.546 & 1.408\\
        & B & Core contraction & 2651 & 8.838  & 3.787 & 0.653 & 1.888 & 3.806 & 0.779 & 2.000\\
        & C & Terminal age main sequence & 2823 & 8.832  & 3.798 & 0.785 & 2.092 & 3.792 & 0.817 & 2.226\\
        & D & Thin H shell burning & 3152 & 8.817 & 3.775 & 0.819 & 2.410 & 3.741 & 0.745 & 2.591\\
        & E & Entering red giant phase & 3351 & 8.800 & 3.713 & 0.661 & 2.665 & 3.695 & 0.857 & 3.648 \\
        & F & Starting of mass transfer & 3541 & 8.678 & 3.691 & 0.940 & 4.079 & 3.657 & 1.526 & 9.382 \\
        \midrule
        \parbox[t]{2mm}{\multirow{6}{*}{\rotatebox[origin=c]{90}{Secondary}}} & a & Zero age main sequence & 0 & 8.876  & 3.821 & 0.525 & 1.392 & 3.824 & 0.546 & 1.408\\
        & b & Core contraction & 2508 & 8.841 & 3.789 & 0.644 & 1.846 & 3.789 & 0.670 & 1.905  \\
        & c & Terminal age main sequence & 2675 & 8.837 & 3.787 & 0.658 & 1.897 & 3.800 & 0.800 & 2.104 \\
        & d & Thin H shell burning & 3028 & 8.824 & 3.786 & 0.810 & 2.271 & 3.776 & 0.838 & 2.452 \\ 
        & e & Entering red giant phase & 3213 & 8.814 & 3.766 & 0.810 & 2.493 & 3.714 & 0.672 & 2.702 \\
        & f & Starting of mass transfer & 3541 & 8.678 & 3.691 & 0.940 & 4.079 & 3.657 & 1.526 & 9.382 \\
        \bottomrule
        \end{tabular}}
    \label{tab:rsaritime}
\end{table*}

\subsubsection{V455 Aur}

\begin{figure}
    \centering
    \includegraphics[width=0.46\textwidth]{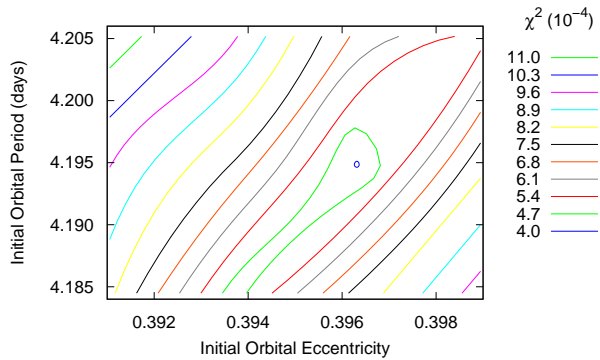}
    \caption{$\chi^2$ map of corresponding initial period for each initial orbital eccentricity of V455 Aur.}
    \label{fig:v455aurchi}
\end{figure}

\begin{figure}
    \centering
    \centering
    \begin{subfigure}
        \centering
        \includegraphics[width=0.45\textwidth]{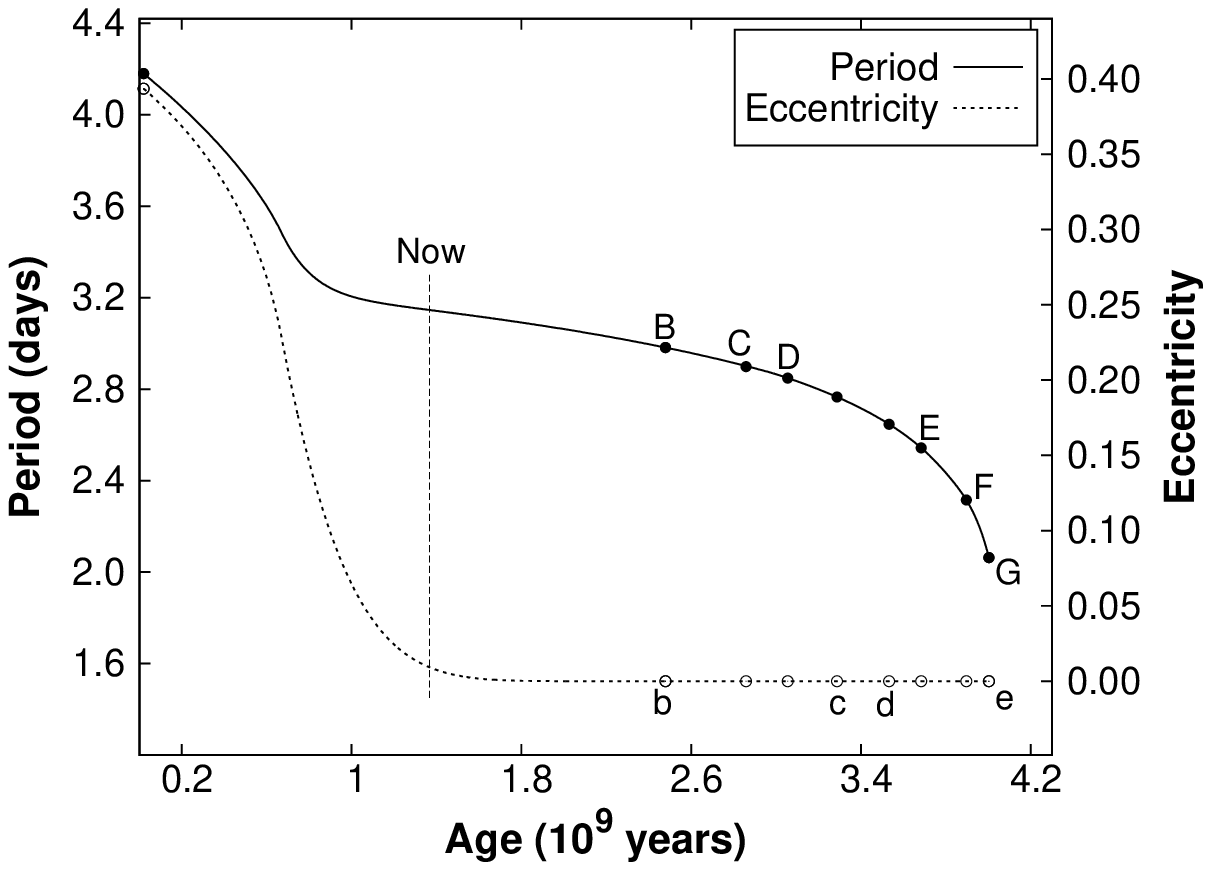}
    \end{subfigure}
    \begin{subfigure}
        \centering
        \includegraphics[width=0.45\textwidth]{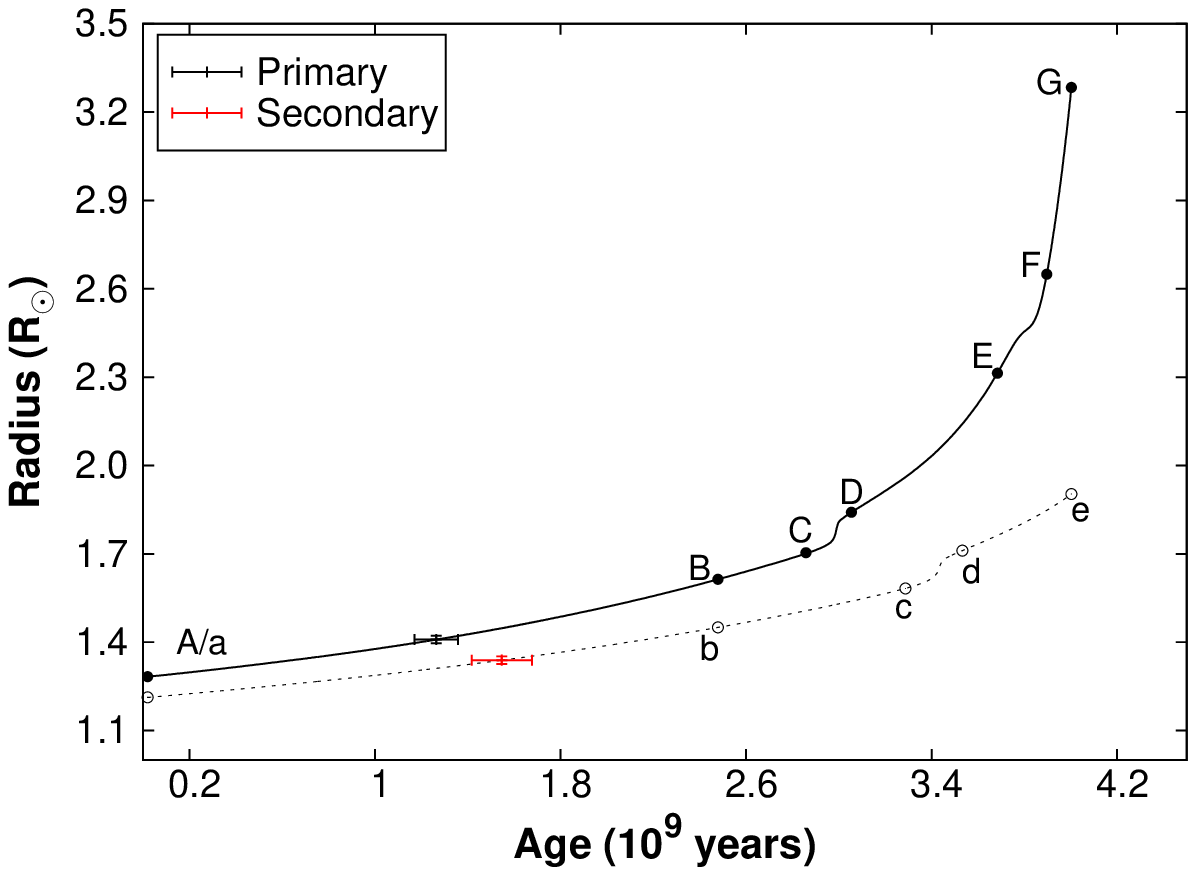}
    \end{subfigure}
    \caption{Change of orbital parameters and radius of components of V455 Aur in time}
    \label{fig:mesav455aur}
\end{figure}

\begin{figure}
    \centering
    \includegraphics[width=0.46\textwidth]{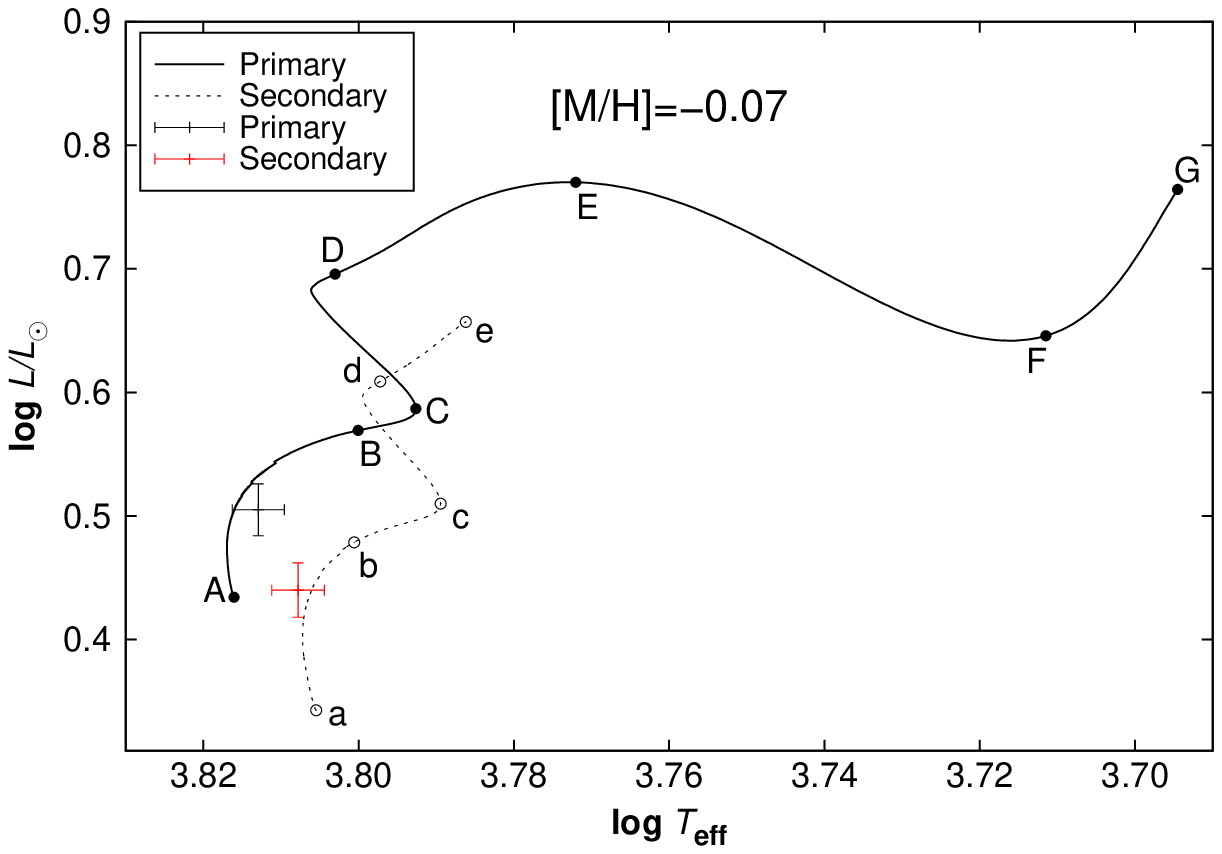}
    \caption{Detailed evolution of V455 Aur on $T_\mathrm{eff}$-log$L/{L_\odot}$ plane}
    \label{fig:v455aurhr}
\end{figure}

Orbit of V455 Aur is almost circular, $e$=0.00930, but regarding evolutionary status, it indicates that the system has started its evolution with an eccentricity. Hence, we used the same procedure, which we used for AN Cam. We built evolution models that initial parameters change for the period and eccentricity between 4.050 and 4.350 days with an interval of 0.005 days and between 0.370 and 0.420 with an interval of 0.002, respectively, and kept the evolution continued until eccentricity drops to 0.00930, which is the up-to-date eccentricity value of V455 Aur. Finally, we selected the best model, which gives the smallest $\chi^2$ value, 0.0003678, with initial orbital parameters for period and eccentricity as, 4.195 and 0.396, respectively (given in Fig. \ref{fig:v455aurchi}). After that, we made an evolution model with those initial orbital parameters and stopped the model until the mass transfer started through the Roche lobe. According to our model, the age of V455 Aur is 1.37$\pm$0.25 Gyrs, and the primary component of V455 Aur will start transferring mass through its Roche lobe when the primary component is becoming a red giant while the secondary component is in the phase of burning thick hydrogen shell around its helium core when the age of the system is roughly at 4 Gyrs. The changes in orbital parameters and radii of the components during evolution are presented in Fig. \ref{fig:mesav455aur}. Detailed evolution of both components of V455 Aur with timetables are given in Table \ref{tab:v455time} and Fig.\ref{fig:v455aurhr}. Similar to AN Cam and RS Ari, V455 Aur will become a WD-Giant binary in the late stages of its evolution. The stellar masses will be $M_{\mathrm{WD}}$=0.26 $M_\odot$ and $M_{\mathrm{Giant}}$=1.64 $M_\odot$ under the mass loss and transfer rate assumptions as taken for AN Cam and RS Ari.

\begin{table*}
\centering
    \caption{Detailed evolution of V455 Aur with time stamps}
    \resizebox{\textwidth}{!}{\begin{tabular}{lccccccccccc}
    \toprule
        & \multirow{2}*{Mark} & \multirow{2}*{Evolutionary Status} & Age & Period & \multirow{2}*{Eccentricity} & \multicolumn{3}{c}{Primary} & \multicolumn{3}{c}{Secondary} \\
        & & & (Myears) & (days) &  & log $T_\mathrm{eff}$ (K) & log $L$ (L$_\odot$)  & Radius (R$_\odot$) & log $T_\mathrm{eff}$ (K) & log $L$ (L$_\odot$)  & Radius (R$_\odot$)\\
        \hline
        \parbox[t]{2mm}{\multirow{8}{*}{\rotatebox[origin=c]{90}{Primary}}} & A & Zero age main sequence & 0 & 4.195 & 0.396 & 3.816 & 0.434 & 1.283 & 3.805 & 0.343 & 1.213 \\
        & B & Circularisation of Orbit & 2479 & 2.982 & 0 & 3.800 & 0.569 & 1.613 & 3.800 & 0.479 & 1.450\\
        & C & Core contraction & 2859 & 2.899 & 0 & 3.793 & 0.587 & 1.704 & 3.795 & 0.492 & 1.507\\
        & D & Terminal age main sequence & 3055 & 2.849 & 0 & 3.803 & 0.696 & 1.841 & 3.792 & 0.497 & 1.539\\
        & E & Thin H shell burning & 3684 & 2.543 & 0 & 3.772 & 0.770 & 2.313 & 3.794 & 0.621 & 1.762 \\ 
        & F & Entering red giant phase & 3897 & 2.315 & 0 & 3.711 & 0.646 & 2.650 & 3.789 & 0.643 & 1.848\\
        & G & Starting of mass transfer & 4002 & 2.064 & 0 & 3.694 & 0.764 & 3.284 & 3.786 & 0.657 & 1.903\\
        \midrule
        \parbox[t]{2mm}{\multirow{5}{*}{\rotatebox[origin=c]{90}{Secondary}}} & a & Zero age main sequence & 0 & 4.195 & 0.396 & 3.816 & 0.434 & 1.283 & 3.805 & 0.343 & 1.213 \\
        & b & Circularisation of Orbit & 2479 & 2.982 & 0 & 3.800 & 0.569 & 1.613 & 3.800 & 0.479 & 1.450\\
        & c & Core contraction & 3287 & 2.766 & 0 & 3.795 & 0.720 & 1.961 & 3.789 & 0.510 & 1.582\\
        & d & Terminal age main sequence & 3533 & 2.647 & 0 & 3.785 & 0.755 & 2.141 & 3.797 & 0.609 & 1.711 \\
        & e & Starting of mass transfer & 4002 & 2.064 & 0 & 3.694 & 0.764 & 3.284 & 3.786 & 0.657 & 1.903\\
        \bottomrule
    \end{tabular}}
    \label{tab:v455time}
\end{table*}

\subsubsection{Uncertainty of Fundamental Stellar Parameters Prior to Mass-Transfer}

The orbital period affects the size of the orbit prior to mass-transfer which determines the fate of a binary system. Due to the uncertainty of each orbital parameter (orbital period, eccentricity), a deviation is expected in the initial orbital parameters used in the {\sc mesa}-code which would affect the final evolutionary parameters. A grid of evolutionary models is calculated with orbital parameters with changing values within their individual uncertainty box to see how initial orbital parameters deviate. It is seen that the deviations in the initial orbital period and orbital eccentricity are within 1 percent and 0.1 percent, respectively. These deviations show themselves in the final fundamental parameters with similar amounts. Therefore, the timestamps  given in Tables~\ref{tab:ancamtime},\ref{tab:rsaritime} and \ref{tab:v455time} are stable within the uncertainty box of the orbital parameters obtained in this study.

\begin{table*}
\centering
\caption{Comparison of results of AN Cam and V455 Aur with \citetalias{Southworth2021} and \citetalias{Southworth2021b}.}
\resizebox{\textwidth}{!}{\begin{tabular}{ccccccccc}
\toprule
\multirow{3}*{Parameters} & \multicolumn{4}{c}{AN Cam} & \multicolumn{4}{c}{V455 Aur} \\
& \multicolumn{2}{c}{This Study} & \multicolumn{2}{c}{\citetalias{Southworth2021}} & \multicolumn{2}{c}{This study} & \multicolumn{2}{c}{\citetalias{Southworth2021b}} \\
& Primary & Secondary & Primary & Secondary & Primary & Secondary & Primary & Secondary \\
\hline
Mass (M$_\odot$) & 1.383$\pm$0.025 & 1.406$\pm$0.024 & 1.380$\pm$0.021 & 1.402$\pm$0.025 & 1.287$\pm$0.003 & 1.231$\pm$0.003 & 1.2887$\pm$0.0063 & 1.2316$\pm$0.0050\\
Radius ($R_\odot$) & 2.206$\pm$0.059 & 2.667$\pm$0.058 & 2.159$\pm$0.012 & 2.646$\pm$0.014 & 1.409$\pm$0.013 & 1.339$\pm$0.013 & 1.389$\pm$0.011 & 1.318$\pm$0.014\\
a (R$_\odot$) & \multicolumn{2}{c}{45.065$\pm$0.252} & \multicolumn{2}{c}{45.05$\pm$0.24} & \multicolumn{2}{c}{12.286$\pm$0.007} & \multicolumn{2}{c}{12.295$\pm$0.017}\\
log g (cgs) & 3.892$\pm$0.059 & 3.734$\pm$0.055 & 3.9095$\pm$0.0030 & 3.7400$\pm$0.0037 & 4.250$\pm$0.009 & 4.275$\pm$0.010 & 4.2626$\pm$0.0070 & 4.2887$\pm$0.0089\\
T$_\mathrm{eff}$ (K) & 6050$\pm$100 & 5900$\pm$100 & 6050$\pm$150 & 5750$\pm$150 & 6500$\pm$50 & 6424$\pm$50 & 6500$\pm$200 & 6400$\pm$200\\
log$L$ (L$_\odot$) & 0.787$\pm$0.036 & 0.891$\pm$0.034 & 0.750$\pm$0.043 & 0.839$\pm$0.046 & 0.505$\pm$0.021 & 0.440$\pm$0.022 & 0.492$\pm$0.054 & 0.419 $\pm$ 0.055\\
\text{[M/H]} & \multicolumn{2}{c}{0.00$\pm$0.12} & \multicolumn{2}{c}{0.00$\pm$0.07} & \multicolumn{2}{c}{-0.07$\pm$0.07} & \multicolumn{2}{c}{-0.07$\pm$0.07} \\
Age (Gyr) & \multicolumn{2}{c}{3.00$\pm$0.15} & \multicolumn{2}{c}{3.3} & \multicolumn{2}{c}{1.37$\pm$0.25} & \multicolumn{2}{c}{1.8$\pm$0.2}\\
\bottomrule
\label{tab:comparison}
\end{tabular}}
\end{table*}

\section{Conclusions}

Twin binary systems (q $>$ 0.95), generally, are interesting objects with their characteristic features. In this study, we have selected three twin binary systems and derived their fundamental parameters, including temperature, metallicity, and age, of three eclipsing binaries, by combining their TESS data, precise RV measurements, and new high resolution spectra. Two of the systems (AN Cam and RS Ari) are found to be evolved, which means they are post terminal age main sequence (T.A.M.S.) systems, while the other one (V455 Aur) is still a main-sequence system. More accurate spectroscopy related parameters of the systems such as metallicity and effective temperature are determined then previous studies on the same objects \citep[][]{Southworth2021, Southworth2021b}. All three systems are found to have nearly Solar metallicity within the uncertainty of measurements. 
Besides, the distances of these three systems we analyzed are in excellent agreement with \textit{Gaia} eDR3\citep[][]{2020gaia} distances, which emphasized our measurements are accurate. The third body in V455 Aur has not been studied in this paper except its light contribution. We obtained a similar light contribution to the one obtained by \cite{Southworth2021}, which will yield similar physical parameters of the third star. Finally, comparison of our results with \citetalias{Southworth2021} and \citetalias{Southworth2021b} for AN Cam and V455 Aur, respectively, is given in Table \ref{tab:comparison}. The primary and secondary components of AN Cam are rotating much faster than their synchronisation velocity, which is expected as the orbit of the system is still not circularized (see Table \ref{tab:ancamtime}). In other two systems, component stars of V455 Aur have synchronised rotation velocities with their orbits as expected from their evolutionary times stamps (see Table \ref{tab:v455time}) while the secondary component of RS Ari is rotating faster than its synchronisation velocity. This is interesting because RS Ari completed the orbital circularization time scale and the primary component has a synchronised rotational velocity. Calculating the synchronisation time scale using \cite{Zahn} for RS Ari yields $\sim$ 61 Myr which is much smaller than RS Ari's present age.

With help of modern evolution code, \textsc{MESA}, we have calculated the evolutionary phases of these three systems including initial orbital parameters, which could shed light on, generally, understanding the properties of current semi-detached binaries. There are still very few systems, in which their evolutionary phases have been revealed. We believe that finding the initial orbital properties of twin binaries would help to understand the formation mechanism of twin binaries in detail.

\section*{Acknowledgements}
We are grateful to the anonymous referee for her/his valuable suggestions. This work is a part of the PhD thesis of GY. We thank to Akdeniz University Space Sciences and Technologies Department for granting us observing time with UBT60 telescope. We would like to thank Dr. A. J. Rosales for helping us in {\sc mesa} calculations. This paper includes data collected with the TESS mission, obtained from the MAST data archive at the Space Telescope Science Institute (STScI). Funding for the {\it TESS} mission is provided by the NASA Explorer Program. STScI is operated by the Association of Universities for Research in Astronomy, Inc., under NASA contract NAS 5–26555. This research has made use of the SIMBAD database, operated at CDS, Strasbourg, France. This work has made use of data from the European Space Agency (ESA) mission {\it Gaia} (\url{https://www.cosmos.esa.int/gaia}), processed by the {\it Gaia} Data Processing and Analysis Consortium (DPAC,
\url{https://www.cosmos.esa.int/web/gaia/dpac/consortium}). Funding for the DPAC has been provided by national institutions, in particular the institutions participating in the {\it Gaia} Multilateral Agreement. This research has made use of NASA’s Astrophysics Data System.

\section*{Data Availability}

All data are incorporated into the article.



\bibliographystyle{mnras}
\bibliography{reference} 

\newcommand{\noop}[1]{}
\begin{thebibliography}{}
\makeatletter
\relax
\def\mn@urlcharsother{\let\do\@makeother \do\$\do\&\do\#\do\^\do\_\do\%\do\~}
\def\mn@doi{\begingroup\mn@urlcharsother \@ifnextchar [ {\mn@doi@}
  {\mn@doi@[]}}
\def\mn@doi@[#1]#2{\def\@tempa{#1}\ifx\@tempa\@empty \href
  {http://dx.doi.org/#2} {doi:#2}\else \href {http://dx.doi.org/#2} {#1}\fi
  \endgroup}
\def\mn@eprint#1#2{\mn@eprint@#1:#2::\@nil}
\def\mn@eprint@arXiv#1{\href {http://arxiv.org/abs/#1} {{\tt arXiv:#1}}}
\def\mn@eprint@dblp#1{\href {http://dblp.uni-trier.de/rec/bibtex/#1.xml}
  {dblp:#1}}
\def\mn@eprint@#1:#2:#3:#4\@nil{\def\@tempa {#1}\def\@tempb {#2}\def\@tempc
  {#3}\ifx \@tempc \@empty \let \@tempc \@tempb \let \@tempb \@tempa \fi \ifx
  \@tempb \@empty \def\@tempb {arXiv}\fi \@ifundefined
  {mn@eprint@\@tempb}{\@tempb:\@tempc}{\expandafter \expandafter \csname
  mn@eprint@\@tempb\endcsname \expandafter{\@tempc}}}

\bibitem[\protect\citeauthoryear{Aitken \& Silverstone}{Aitken \&
  Silverstone}{1942}]{aitken1942xv}
Aitken A.,  Silverstone H.,  1942, Proceedings of the Royal Society of
  Edinburgh Section A: Mathematics, 61, 186

\bibitem[\protect\citeauthoryear{Albrecht et~al.,}{Albrecht
  et~al.}{2009}]{albrecht2009findings}
Albrecht A.,  et~al., 2009, arXiv preprint arXiv:0901.0721

\bibitem[\protect\citeauthoryear{{Andersen}}{{Andersen}}{1991}]{Andersen}
{Andersen} J.,  1991, \mn@doi [\aapr] {10.1007/BF00873538}, \href
  {https://ui.adsabs.harvard.edu/abs/1991A&ARv...3...91A} {3, 91}

\bibitem[\protect\citeauthoryear{{Bak{\i}{\c{s}}}, {Bak{\i}{\c{s}}}, {Demircan}
   \& {Eker}}{{Bak{\i}{\c{s}}} et~al.}{2008}]{HBakis2008}
{Bak{\i}{\c{s}}} H.,  {Bak{\i}{\c{s}}} V.,  {Demircan} O.,   {Eker} Z.,  2008,
  \mn@doi [\mnras] {10.1111/j.1365-2966.2007.12832.x}, \href
  {https://ui.adsabs.harvard.edu/abs/2008MNRAS.385..381B} {385, 381}

\bibitem[\protect\citeauthoryear{{Bak{\i}{\c{s}}}, {Bak{\i}{\c{s}}}, {Eker}  \&
  {Demircan}}{{Bak{\i}{\c{s}}} et~al.}{2016}]{HBakis2016}
{Bak{\i}{\c{s}}} H.,  {Bak{\i}{\c{s}}} V.,  {Eker} Z.,   {Demircan} O.,  2016,
  \mn@doi [\mnras] {10.1093/mnras/stw320}, \href
  {https://ui.adsabs.harvard.edu/abs/2016MNRAS.458..508B} {458, 508}

\bibitem[\protect\citeauthoryear{{Bak{\i}{\c{s}}}, {Eker}, {Sar{\i}},
  {Y{\"u}cel}  \& {Sonba{\c{s}}}}{{Bak{\i}{\c{s}}} et~al.}{2020}]{Bakis2020}
{Bak{\i}{\c{s}}} V.,  {Eker} Z.,  {Sar{\i}} O.,  {Y{\"u}cel} G.,
  {Sonba{\c{s}}} E.,  2020, \mn@doi [\mnras] {10.1093/mnras/staa1587}, \href
  {https://ui.adsabs.harvard.edu/abs/2020MNRAS.496.2605B} {496, 2605}

\bibitem[\protect\citeauthoryear{{Bak{\i}{\c{s}}}, {K{\"o}seoglu},
  {Bak{\i}{\c{s}}}, {Nitschelm}  \& {Eker}}{{Bak{\i}{\c{s}}}
  et~al.}{2021}]{HBakis2021}
{Bak{\i}{\c{s}}} H.,  {K{\"o}seoglu} D.~T.,  {Bak{\i}{\c{s}}} V.,  {Nitschelm}
  C.,   {Eker} Z.,  2021, \mn@doi [\mnras] {10.1093/mnras/stab560}, \href
  {https://ui.adsabs.harvard.edu/abs/2021MNRAS.503.2432B} {503, 2432}

\bibitem[\protect\citeauthoryear{{Baraffe}, {Chabrier}, {Allard}  \&
  {Hauschildt}}{{Baraffe} et~al.}{1998}]{Baraffe}
{Baraffe} I.,  {Chabrier} G.,  {Allard} F.,   {Hauschildt} P.~H.,  1998, \aap,
  \href {https://ui.adsabs.harvard.edu/abs/1998A&A...337..403B} {337, 403}

\bibitem[\protect\citeauthoryear{{Baranne}, {Mayor}  \& {Poncet}}{{Baranne}
  et~al.}{1979}]{Baranne1979}
{Baranne} A.,  {Mayor} M.,   {Poncet} J.~L.,  1979, \mn@doi [Vistas in
  Astronomy] {10.1016/0083-6656(79)90016-3}, \href
  {https://ui.adsabs.harvard.edu/abs/1979VA.....23..279B} {23, 279}

\bibitem[\protect\citeauthoryear{{Bate}}{{Bate}}{2000}]{Bate2000}
{Bate} M.~R.,  2000, \mn@doi [\mnras] {10.1046/j.1365-8711.2000.03333.x}, \href
  {https://ui.adsabs.harvard.edu/abs/2000MNRAS.314...33B} {314, 33}

\bibitem[\protect\citeauthoryear{{Bate}}{{Bate}}{2019}]{Bate2019}
{Bate} M.~R.,  2019, \mn@doi [\mnras] {10.1093/mnras/stz103}, \href
  {https://ui.adsabs.harvard.edu/abs/2019MNRAS.484.2341B} {484, 2341}

\bibitem[\protect\citeauthoryear{{Bate}, {Bonnell}  \& {Bromm}}{{Bate}
  et~al.}{2002}]{Bate2002}
{Bate} M.~R.,  {Bonnell} I.~A.,   {Bromm} V.,  2002, \mn@doi [\mnras]
  {10.1046/j.1365-8711.2002.05775.x}, \href
  {https://ui.adsabs.harvard.edu/abs/2002MNRAS.336..705B} {336, 705}

\bibitem[\protect\citeauthoryear{{Benedict} et~al.,}{{Benedict}
  et~al.}{2016}]{Benedict}
{Benedict} G.~F.,  et~al., 2016, \mn@doi [\aj] {10.3847/0004-6256/152/5/141},
  \href {https://ui.adsabs.harvard.edu/abs/2016AJ....152..141B} {152, 141}

\bibitem[\protect\citeauthoryear{{Borucki} et~al.,}{{Borucki}
  et~al.}{2010}]{Borucki}
{Borucki} W.~J.,  et~al., 2010, \mn@doi [Science] {10.1126/science.1185402},
  \href {https://ui.adsabs.harvard.edu/abs/2010Sci...327..977B} {327, 977}

\bibitem[\protect\citeauthoryear{{Bressan}, {Marigo}, {Girardi}, {Salasnich},
  {Dal Cero}, {Rubele}  \& {Nanni}}{{Bressan} et~al.}{2012}]{Bressan}
{Bressan} A.,  {Marigo} P.,  {Girardi} L.,  {Salasnich} B.,  {Dal Cero} C.,
  {Rubele} S.,   {Nanni} A.,  2012, \mn@doi [\mnras]
  {10.1111/j.1365-2966.2012.21948.x}, \href
  {https://ui.adsabs.harvard.edu/abs/2012MNRAS.427..127B} {427, 127}

\bibitem[\protect\citeauthoryear{{Budding}, {Bakis}, {Erdem}, {Demircan},
  {Iliev}, {Iliev}  \& {Slee}}{{Budding} et~al.}{2005}]{Budding2005}
{Budding} E.,  {Bakis} V.,  {Erdem} A.,  {Demircan} O.,  {Iliev} L.,  {Iliev}
  I.,   {Slee} O.~B.,  2005, \mn@doi [\apss] {10.1007/s10509-005-4855-7}, \href
  {https://ui.adsabs.harvard.edu/abs/2005Ap&SS.296..371B} {296, 371}

\bibitem[\protect\citeauthoryear{{Chabrier}}{{Chabrier}}{2003}]{Chabrier}
{Chabrier} G.,  2003, \mn@doi [\apjl] {10.1086/374879}, \href
  {https://ui.adsabs.harvard.edu/abs/2003ApJ...586L.133C} {586, L133}

\bibitem[\protect\citeauthoryear{{Demircan} \& {Kahraman}}{{Demircan} \&
  {Kahraman}}{1991}]{Demircan}
{Demircan} O.,  {Kahraman} G.,  1991, \mn@doi [\apss] {10.1007/BF00639097},
  \href {https://ui.adsabs.harvard.edu/abs/1991Ap&SS.181..313D} {181, 313}

\bibitem[\protect\citeauthoryear{{Duquennoy} \& {Mayor}}{{Duquennoy} \&
  {Mayor}}{1991}]{Duquennoy}
{Duquennoy} A.,  {Mayor} M.,  1991, \aap, \href
  {https://ui.adsabs.harvard.edu/abs/1991A&A...248..485D} {500, 337}

\bibitem[\protect\citeauthoryear{{Eker} et~al.,}{{Eker} et~al.}{2018}]{Eker}
{Eker} Z.,  et~al., 2018, \mn@doi [\mnras] {10.1093/mnras/sty1834}, \href
  {https://ui.adsabs.harvard.edu/abs/2018MNRAS.479.5491E} {479, 5491}

\bibitem[\protect\citeauthoryear{{Eker} et~al.,}{{Eker}
  et~al.}{2020}]{2020eker}
{Eker} Z.,  et~al., 2020, \mn@doi [\mnras] {10.1093/mnras/staa1659}, \href
  {https://ui.adsabs.harvard.edu/abs/2020MNRAS.496.3887E} {496, 3887}

\bibitem[\protect\citeauthoryear{{El-Badry}, {Rix}, {Tian}, {Duch{\^e}ne}  \&
  {Moe}}{{El-Badry} et~al.}{2019}]{Elbadry2019}
{El-Badry} K.,  {Rix} H.-W.,  {Tian} H.,  {Duch{\^e}ne} G.,   {Moe} M.,  2019,
  \mn@doi [\mnras] {10.1093/mnras/stz2480}, \href
  {https://ui.adsabs.harvard.edu/abs/2019MNRAS.489.5822E} {489, 5822}

\bibitem[\protect\citeauthoryear{Fr{\'e}chet}{Fr{\'e}chet}{1943}]{frechet1943}
Fr{\'e}chet M.,  1943, Revue de l'Institut International de Statistique, pp
  182--205

\bibitem[\protect\citeauthoryear{{Gaia Collaboration}}{{Gaia
  Collaboration}}{2020}]{2020gaia}
{Gaia Collaboration} 2020, VizieR Online Data Catalog, \href
  {https://ui.adsabs.harvard.edu/abs/2020yCat.1350....0G} {p. I/350}

\bibitem[\protect\citeauthoryear{{Gray} \& {Corbally}}{{Gray} \&
  {Corbally}}{1994}]{Gray}
{Gray} R.~O.,  {Corbally} C.~J.,  1994, \mn@doi [\aj] {10.1086/116893}, \href
  {https://ui.adsabs.harvard.edu/abs/1994AJ....107..742G} {107, 742}

\bibitem[\protect\citeauthoryear{{Griffin}}{{Griffin}}{1967}]{Griffin1967}
{Griffin} R.~F.,  1967, \mn@doi [\apj] {10.1086/149168}, \href
  {https://ui.adsabs.harvard.edu/abs/1967ApJ...148..465G} {148, 465}

\bibitem[\protect\citeauthoryear{{Griffin}}{{Griffin}}{2001}]{Griffin2001}
{Griffin} R.~F.,  2001, The Observatory, \href
  {https://ui.adsabs.harvard.edu/abs/2001Obs...121..315G} {121, 315}

\bibitem[\protect\citeauthoryear{{Griffin}}{{Griffin}}{2013}]{Griffin2013}
{Griffin} R.~F.,  2013, The Observatory, \href
  {https://ui.adsabs.harvard.edu/abs/2013Obs...133..156G} {133, 156}

\bibitem[\protect\citeauthoryear{{Halbwachs}, {Mayor}, {Udry}  \&
  {Arenou}}{{Halbwachs} et~al.}{2003}]{Halb2003}
{Halbwachs} J.~L.,  {Mayor} M.,  {Udry} S.,   {Arenou} F.,  2003, \mn@doi
  [\aap] {10.1051/0004-6361:20021507}, \href
  {https://ui.adsabs.harvard.edu/abs/2003A&A...397..159H} {397, 159}

\bibitem[\protect\citeauthoryear{{H{\"o}fner} \& {Olofsson}}{{H{\"o}fner} \&
  {Olofsson}}{2018}]{Hofner2018}
{H{\"o}fner} S.,  {Olofsson} H.,  2018, \mn@doi [\aapr]
  {10.1007/s00159-017-0106-5}, \href
  {https://ui.adsabs.harvard.edu/abs/2018A&ARv..26....1H} {26, 1}

\bibitem[\protect\citeauthoryear{{H{\o}g} et~al.,}{{H{\o}g} et~al.}{2000}]{Hog}
{H{\o}g} E.,  et~al., 2000, \aap, \href
  {https://ui.adsabs.harvard.edu/abs/2000A&A...355L..27H} {355, L27}

\bibitem[\protect\citeauthoryear{{Hurley}, {Tout}  \& {Pols}}{{Hurley}
  et~al.}{2002}]{Hurley}
{Hurley} J.~R.,  {Tout} C.~A.,   {Pols} O.~R.,  2002, \mn@doi [\mnras]
  {10.1046/j.1365-8711.2002.05038.x}, \href
  {https://ui.adsabs.harvard.edu/abs/2002MNRAS.329..897H} {329, 897}

\bibitem[\protect\citeauthoryear{{Imbert}}{{Imbert}}{1987}]{imbert1987}
{Imbert} M.,  1987, \aaps, \href
  {https://ui.adsabs.harvard.edu/abs/1987A&AS...67..161I} {67, 161}

\bibitem[\protect\citeauthoryear{{Imbert}}{{Imbert}}{2002}]{imbert2002}
{Imbert} M.,  2002, \mn@doi [\aap] {10.1051/0004-6361:20020392}, \href
  {https://ui.adsabs.harvard.edu/abs/2002A&A...387..850I} {387, 850}

\bibitem[\protect\citeauthoryear{{Kounkel} et~al.,}{{Kounkel}
  et~al.}{2019}]{Kounkel2019}
{Kounkel} M.,  et~al., 2019, \mn@doi [\aj] {10.3847/1538-3881/ab13b1}, \href
  {https://ui.adsabs.harvard.edu/abs/2019AJ....157..196K} {157, 196}

\bibitem[\protect\citeauthoryear{{Kurucz}}{{Kurucz}}{1979}]{Kurucz1979}
{Kurucz} R.~L.,  1979, \mn@doi [\apjs] {10.1086/190589}, \href
  {https://ui.adsabs.harvard.edu/abs/1979ApJS...40....1K} {40, 1}

\bibitem[\protect\citeauthoryear{{Lucy}}{{Lucy}}{2006}]{Lucy2006}
{Lucy} L.~B.,  2006, \mn@doi [\aap] {10.1051/0004-6361:20065746}, \href
  {https://ui.adsabs.harvard.edu/abs/2006A&A...457..629L} {457, 629}

\bibitem[\protect\citeauthoryear{{Maxted} et~al.,}{{Maxted}
  et~al.}{2020}]{Maxted2020}
{Maxted} P.~F.~L.,  et~al., 2020, \mn@doi [\mnras] {10.1093/mnras/staa1662},
  \href {https://ui.adsabs.harvard.edu/abs/2020MNRAS.498..332M} {498, 332}

\bibitem[\protect\citeauthoryear{{Moe} \& {Di Stefano}}{{Moe} \& {Di
  Stefano}}{2017}]{Moe}
{Moe} M.,  {Di Stefano} R.,  2017, \mn@doi [\apjs] {10.3847/1538-4365/aa6fb6},
  \href {https://ui.adsabs.harvard.edu/abs/2017ApJS..230...15M} {230, 15}

\bibitem[\protect\citeauthoryear{{Oelkers} \& {Stassun}}{{Oelkers} \&
  {Stassun}}{2018}]{Oelkers2018}
{Oelkers} R.~J.,  {Stassun} K.~G.,  2018, \mn@doi [\aj]
  {10.3847/1538-3881/aad68e}, \href
  {https://ui.adsabs.harvard.edu/abs/2018AJ....156..132O} {156, 132}

\bibitem[\protect\citeauthoryear{{Paxton}, {Bildsten}, {Dotter}, {Herwig},
  {Lesaffre}  \& {Timmes}}{{Paxton} et~al.}{2011}]{Paxton2011}
{Paxton} B.,  {Bildsten} L.,  {Dotter} A.,  {Herwig} F.,  {Lesaffre} P.,
  {Timmes} F.,  2011, \mn@doi [\apjs] {10.1088/0067-0049/192/1/3}, \href
  {https://ui.adsabs.harvard.edu/abs/2011ApJS..192....3P} {192, 3}

\bibitem[\protect\citeauthoryear{{Paxton} et~al.,}{{Paxton}
  et~al.}{2013}]{Paxton2013}
{Paxton} B.,  et~al., 2013, \mn@doi [\apjs] {10.1088/0067-0049/208/1/4}, \href
  {https://ui.adsabs.harvard.edu/abs/2013ApJS..208....4P} {208, 4}

\bibitem[\protect\citeauthoryear{{Paxton} et~al.,}{{Paxton}
  et~al.}{2015}]{Paxton2015}
{Paxton} B.,  et~al., 2015, \mn@doi [\apjs] {10.1088/0067-0049/220/1/15}, \href
  {https://ui.adsabs.harvard.edu/abs/2015ApJS..220...15P} {220, 15}

\bibitem[\protect\citeauthoryear{{Paxton} et~al.,}{{Paxton}
  et~al.}{2018}]{Paxton2018}
{Paxton} B.,  et~al., 2018, \mn@doi [\apjs] {10.3847/1538-4365/aaa5a8}, \href
  {https://ui.adsabs.harvard.edu/abs/2018ApJS..234...34P} {234, 34}

\bibitem[\protect\citeauthoryear{{Paxton} et~al.,}{{Paxton}
  et~al.}{2019}]{Paxton2019}
{Paxton} B.,  et~al., 2019, \mn@doi [\apjs] {10.3847/1538-4365/ab2241}, \href
  {https://ui.adsabs.harvard.edu/abs/2019ApJS..243...10P} {243, 10}

\bibitem[\protect\citeauthoryear{Perryman et~al.}{Perryman
  et~al.}{1997}]{Hipparcos}
Perryman M.,  et~al., 1997, Space Astrometry Mission (ESA SP-1200

\bibitem[\protect\citeauthoryear{{Pietrinferni}, {Cassisi}, {Salaris}  \&
  {Castelli}}{{Pietrinferni} et~al.}{2004}]{Pietrinferni}
{Pietrinferni} A.,  {Cassisi} S.,  {Salaris} M.,   {Castelli} F.,  2004,
  \mn@doi [\apj] {10.1086/422498}, \href
  {https://ui.adsabs.harvard.edu/abs/2004ApJ...612..168P} {612, 168}

\bibitem[\protect\citeauthoryear{{Pojmanski}}{{Pojmanski}}{1997}]{Pojmanski}
{Pojmanski} G.,  1997, \actaa, \href
  {https://ui.adsabs.harvard.edu/abs/1997AcA....47..467P} {47, 467}

\bibitem[\protect\citeauthoryear{{Pourbaix} et~al.,}{{Pourbaix}
  et~al.}{2004}]{Pourbaix2004}
{Pourbaix} D.,  et~al., 2004, \mn@doi [\aap] {10.1051/0004-6361:20041213},
  \href {https://ui.adsabs.harvard.edu/abs/2004A&A...424..727P} {424, 727}

\bibitem[\protect\citeauthoryear{{Pr{\v{s}}a} et~al.,}{{Pr{\v{s}}a}
  et~al.}{2011}]{Prsa2011}
{Pr{\v{s}}a} A.,  et~al., 2011, \mn@doi [\aj] {10.1088/0004-6256/141/3/83},
  \href {https://ui.adsabs.harvard.edu/abs/2011AJ....141...83P} {141, 83}

\bibitem[\protect\citeauthoryear{{Pr{\v{s}}a} et~al.,}{{Pr{\v{s}}a}
  et~al.}{2022}]{Prsa2022}
{Pr{\v{s}}a} A.,  et~al., 2022, \mn@doi [\apjs] {10.3847/1538-4365/ac324a},
  \href {https://ui.adsabs.harvard.edu/abs/2022ApJS..258...16P} {258, 16}

\bibitem[\protect\citeauthoryear{{Rappaport}, {Verbunt}  \& {Joss}}{{Rappaport}
  et~al.}{1983}]{Rappaport}
{Rappaport} S.,  {Verbunt} F.,   {Joss} P.~C.,  1983, \mn@doi [\apj]
  {10.1086/161569}, \href
  {https://ui.adsabs.harvard.edu/abs/1983ApJ...275..713R} {275, 713}

\bibitem[\protect\citeauthoryear{{Renzini} \& {Buzzoni}}{{Renzini} \&
  {Buzzoni}}{1986}]{Renzini}
{Renzini} A.,  {Buzzoni} A.,  1986, in {Chiosi} C.,  {Renzini} A.,  eds,
  Astrophysics and Space Science Library Vol. 122, Spectral Evolution of
  Galaxies. pp 195--231, \mn@doi{10.1007/978-94-009-4598-2\_19}

\bibitem[\protect\citeauthoryear{{Ricker} et~al.,}{{Ricker}
  et~al.}{2015}]{TESS}
{Ricker} G.~R.,  et~al., 2015, \mn@doi [Journal of Astronomical Telescopes,
  Instruments, and Systems] {10.1117/1.JATIS.1.1.014003}, \href
  {https://ui.adsabs.harvard.edu/abs/2015JATIS...1a4003R} {1, 014003}

\bibitem[\protect\citeauthoryear{{Rosales}, {Mennickent}, {Schleicher}  \&
  {Senhadji}}{{Rosales} et~al.}{2019}]{Rosales}
{Rosales} J.~A.,  {Mennickent} R.~E.,  {Schleicher} D.~R.~G.,   {Senhadji}
  A.~A.,  2019, \mn@doi [\mnras] {10.1093/mnras/sty3117}, \href
  {https://ui.adsabs.harvard.edu/abs/2019MNRAS.483..862R} {483, 862}

\bibitem[\protect\citeauthoryear{{Sbordone}, {Bonifacio}, {Castelli}  \&
  {Kurucz}}{{Sbordone} et~al.}{2004}]{Castelli}
{Sbordone} L.,  {Bonifacio} P.,  {Castelli} F.,   {Kurucz} R.~L.,  2004,
  Memorie della Societa Astronomica Italiana Supplementi, \href
  {https://ui.adsabs.harvard.edu/abs/2004MSAIS...5...93S} {5, 93}

\bibitem[\protect\citeauthoryear{{Serenelli} et~al.,}{{Serenelli}
  et~al.}{2021}]{Serenelli}
{Serenelli} A.,  et~al., 2021, \mn@doi [\aapr] {10.1007/s00159-021-00132-9},
  \href {https://ui.adsabs.harvard.edu/abs/2021A&ARv..29....4S} {29, 4}

\bibitem[\protect\citeauthoryear{{Shu}, {Adams}  \& {Lizano}}{{Shu}
  et~al.}{1987}]{Shu}
{Shu} F.~H.,  {Adams} F.~C.,   {Lizano} S.,  1987, \mn@doi [\araa]
  {10.1146/annurev.aa.25.090187.000323}, \href
  {https://ui.adsabs.harvard.edu/abs/1987ARA&A..25...23S} {25, 23}

\bibitem[\protect\citeauthoryear{{Simon} \& {Obbie}}{{Simon} \&
  {Obbie}}{2009}]{Simon2009}
{Simon} M.,  {Obbie} R.~C.,  2009, \mn@doi [\aj]
  {10.1088/0004-6256/137/2/3442}, \href
  {https://ui.adsabs.harvard.edu/abs/2009AJ....137.3442S} {137, 3442}

\bibitem[\protect\citeauthoryear{{Southworth}}{{Southworth}}{2021a}]{Southworth2021}
{Southworth} J.,  2021a, The Observatory, \href
  {https://ui.adsabs.harvard.edu/abs/2021Obs...141..122S} {141, 122}

\bibitem[\protect\citeauthoryear{{Southworth}}{{Southworth}}{2021b}]{Southworth2021b}
{Southworth} J.,  2021b, The Observatory, \href
  {https://ui.adsabs.harvard.edu/abs/2021Obs...141..190S} {141, 190}

\bibitem[\protect\citeauthoryear{{Soydugan}, {Soydugan}  \&
  {Ali{\c{c}}avu{\c{s}}}}{{Soydugan} et~al.}{2020}]{Soydugan}
{Soydugan} F.,  {Soydugan} E.,   {Ali{\c{c}}avu{\c{s}}} F.,  2020, \mn@doi
  [Research in Astronomy and Astrophysics] {10.1088/1674-4527/20/4/52}, \href
  {https://ui.adsabs.harvard.edu/abs/2020RAA....20...52S} {20, 052}

\bibitem[\protect\citeauthoryear{{Stassun}}{{Stassun}}{2019}]{Stassun2019}
{Stassun} K.~G.,  2019, VizieR Online Data Catalog, \href
  {https://ui.adsabs.harvard.edu/abs/2019yCat.4038....0S} {p. IV/38}

\bibitem[\protect\citeauthoryear{{Tody}}{{Tody}}{1993}]{Tody}
{Tody} D.,  1993, in {Hanisch} R.~J.,  {Brissenden} R.~J.~V.,   {Barnes} J.,
  eds,  Astronomical Society of the Pacific Conference Series Vol. 52,
  Astronomical Data Analysis Software and Systems II. p.~173

\bibitem[\protect\citeauthoryear{{Tokovinin}}{{Tokovinin}}{2000}]{Tokovinin2000}
{Tokovinin} A.~A.,  2000, \aap, \href
  {https://ui.adsabs.harvard.edu/abs/2000A&A...360..997T} {360, 997}

\bibitem[\protect\citeauthoryear{{Torres}, {Andersen}  \&
  {Gim{\'e}nez}}{{Torres} et~al.}{2010}]{Torres}
{Torres} G.,  {Andersen} J.,   {Gim{\'e}nez} A.,  2010, \mn@doi [\aapr]
  {10.1007/s00159-009-0025-1}, \href
  {https://ui.adsabs.harvard.edu/abs/2010A&ARv..18...67T} {18, 67}

\bibitem[\protect\citeauthoryear{{Udalski}}{{Udalski}}{2003}]{Udalski}
{Udalski} A.,  2003, \actaa, \href
  {https://ui.adsabs.harvard.edu/abs/2003AcA....53..291U} {53, 291}

\bibitem[\protect\citeauthoryear{Van~Hamme}{Van~Hamme}{1993}]{vanhamme1993}
Van~Hamme W.,  1993, \mn@doi [\aj] {10.1086/116788}, \href
  {https://ui.adsabs.harvard.edu/abs/1993AJ....106.2096V} {106, 2096}

\bibitem[\protect\citeauthoryear{{Wilson}}{{Wilson}}{1979}]{1979ApJ...234.1054W}
{Wilson} R.~E.,  1979, \mn@doi [\apj] {10.1086/157588}, \href
  {https://ui.adsabs.harvard.edu/abs/1979ApJ...234.1054W} {234, 1054}

\bibitem[\protect\citeauthoryear{{Wilson}}{{Wilson}}{1990}]{1990ApJ...356..613W}
{Wilson} R.~E.,  1990, \mn@doi [\apj] {10.1086/168867}, \href
  {https://ui.adsabs.harvard.edu/abs/1990ApJ...356..613W} {356, 613}

\bibitem[\protect\citeauthoryear{{Wilson}}{{Wilson}}{2008}]{2008ApJ...672..575W}
{Wilson} R.~E.,  2008, \mn@doi [\apj] {10.1086/523634}, \href
  {https://ui.adsabs.harvard.edu/abs/2008ApJ...672..575W} {672, 575}

\bibitem[\protect\citeauthoryear{{Wilson} \& {Devinney}}{{Wilson} \&
  {Devinney}}{1971}]{1971ApJ...166..605W}
{Wilson} R.~E.,  {Devinney} E.~J.,  1971, \mn@doi [\apj] {10.1086/150986},
  \href {https://ui.adsabs.harvard.edu/abs/1971ApJ...166..605W} {166, 605}

\bibitem[\protect\citeauthoryear{{Wilson} \& {Van Hamme}}{{Wilson} \& {Van
  Hamme}}{2014}]{2014ApJ...780..151W}
{Wilson} R.~E.,  {Van Hamme} W.,  2014, \mn@doi [\apj]
  {10.1088/0004-637X/780/2/151}, \href
  {https://ui.adsabs.harvard.edu/abs/2014ApJ...780..151W} {780, 151}

\bibitem[\protect\citeauthoryear{{Wilson}, {Van Hamme}  \& {Terrell}}{{Wilson}
  et~al.}{2010}]{2010ApJ...723.1469W}
{Wilson} R.~E.,  {Van Hamme} W.,   {Terrell} D.,  2010, \mn@doi [\apj]
  {10.1088/0004-637X/723/2/1469}, \href
  {https://ui.adsabs.harvard.edu/abs/2010ApJ...723.1469W} {723, 1469}

\bibitem[\protect\citeauthoryear{{Y{\"u}cel} \& {Bak{\i}{\c{s}}}}{{Y{\"u}cel}
  \& {Bak{\i}{\c{s}}}}{2022}]{Yucel2022}
{Y{\"u}cel} G.,  {Bak{\i}{\c{s}}} V.,  2022, \mn@doi [\mnras]
  {10.1093/mnras/stac1361}, \href
  {https://ui.adsabs.harvard.edu/abs/2022MNRAS.514...34Y} {514, 34}

\bibitem[\protect\citeauthoryear{{Zahn}}{{Zahn}}{1977}]{Zahn}
{Zahn} J.~P.,  1977, \aap, \href
  {https://ui.adsabs.harvard.edu/abs/1977A&A....57..383Z} {57, 383}

\bibitem[\protect\citeauthoryear{{Zhang}, {Qian}, {Wang}, {Wu}  \&
  {Jiang}}{{Zhang} et~al.}{2017}]{Zhang2017}
{Zhang} J.,  {Qian} S.-B.,  {Wang} S.-M.,  {Wu} Y.,   {Jiang} L.-Q.,  2017,
  \mn@doi [\pasj] {10.1093/pasj/psx023}, \href
  {https://ui.adsabs.harvard.edu/abs/2017PASJ...69...49Z} {69, 49}

\makeatother
\end{thebibliography}




\appendix

\section{Selected Spectral Regions for Metallicity Analysis}

\begin{figure*}
    \centering
    \centering
    \begin{subfigure}
        \centering
        \includegraphics[width=0.45\textwidth]{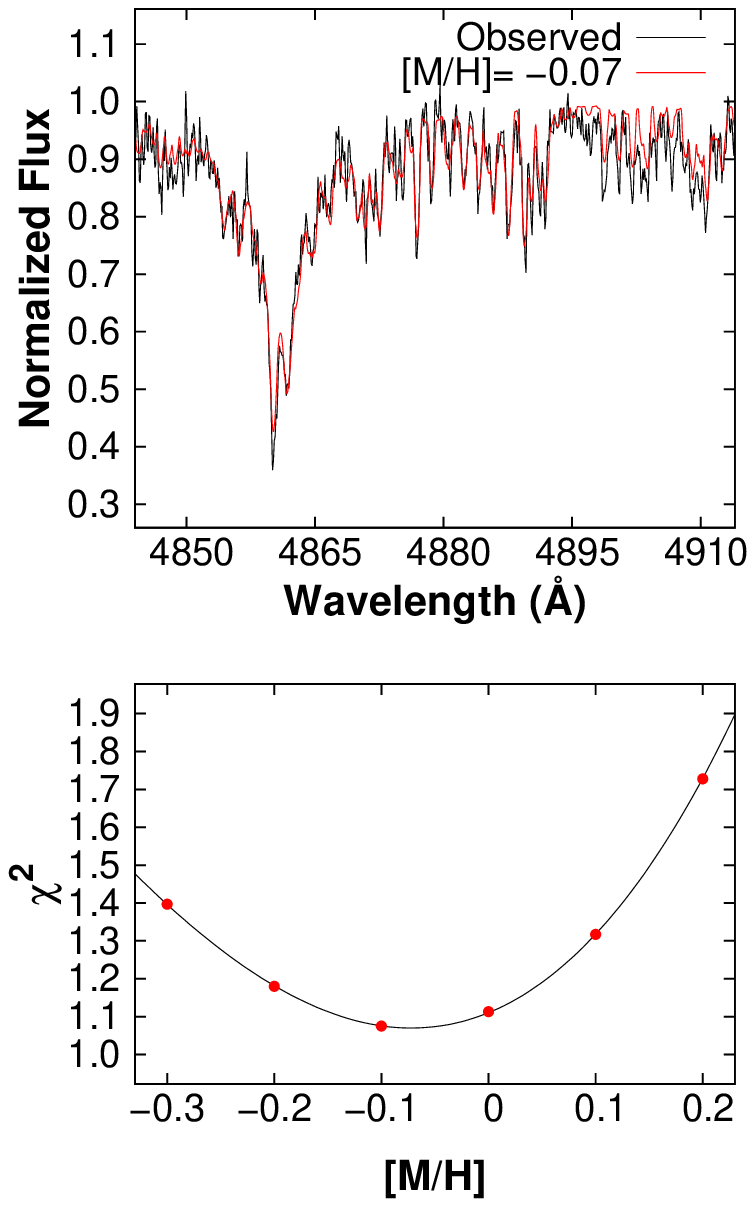}
    \end{subfigure}
    \begin{subfigure}
        \centering
        \includegraphics[width=0.45\textwidth]{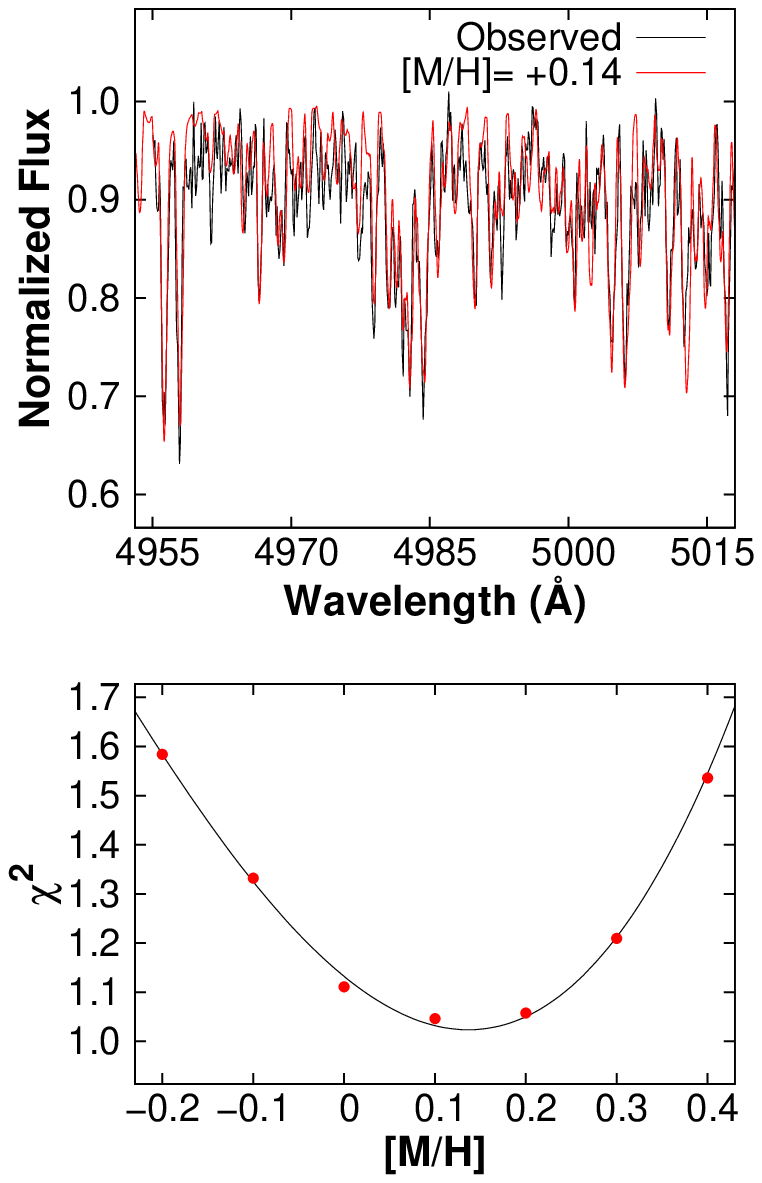}
    \end{subfigure}
    \begin{subfigure}
        \centering
        \includegraphics[width=0.3\textwidth]{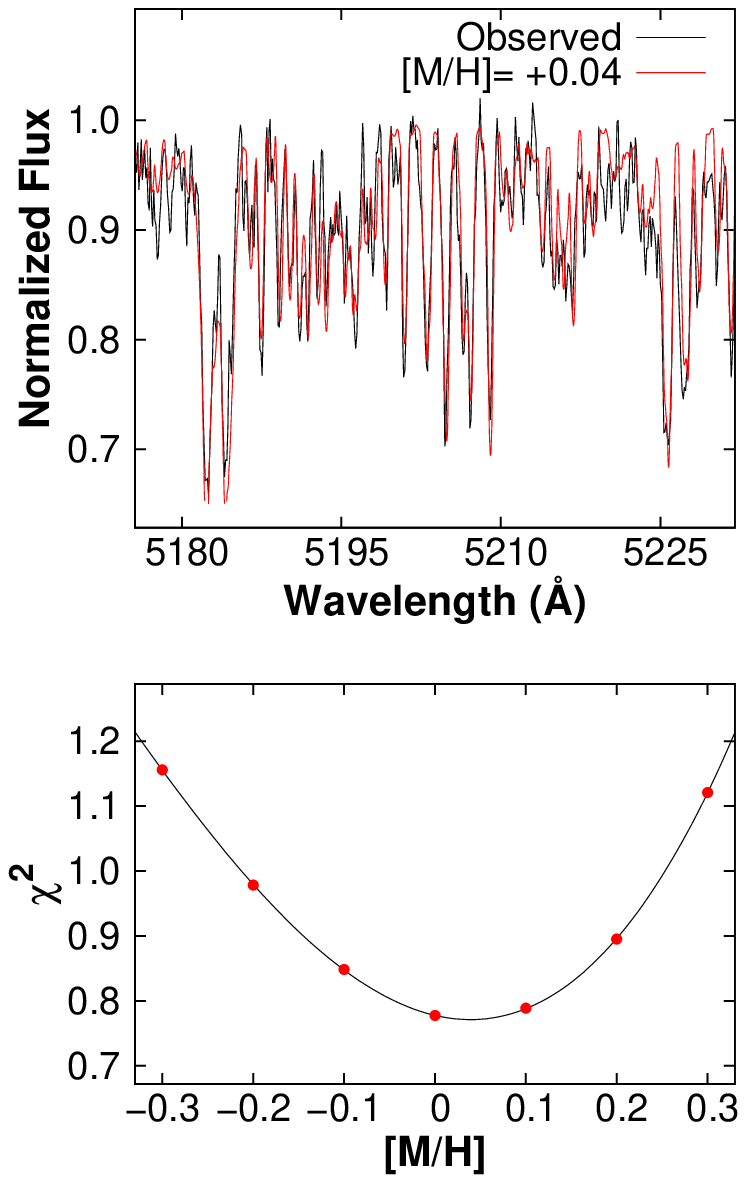}
    \end{subfigure}
    \begin{subfigure}
        \centering
        \includegraphics[width=0.3\textwidth]{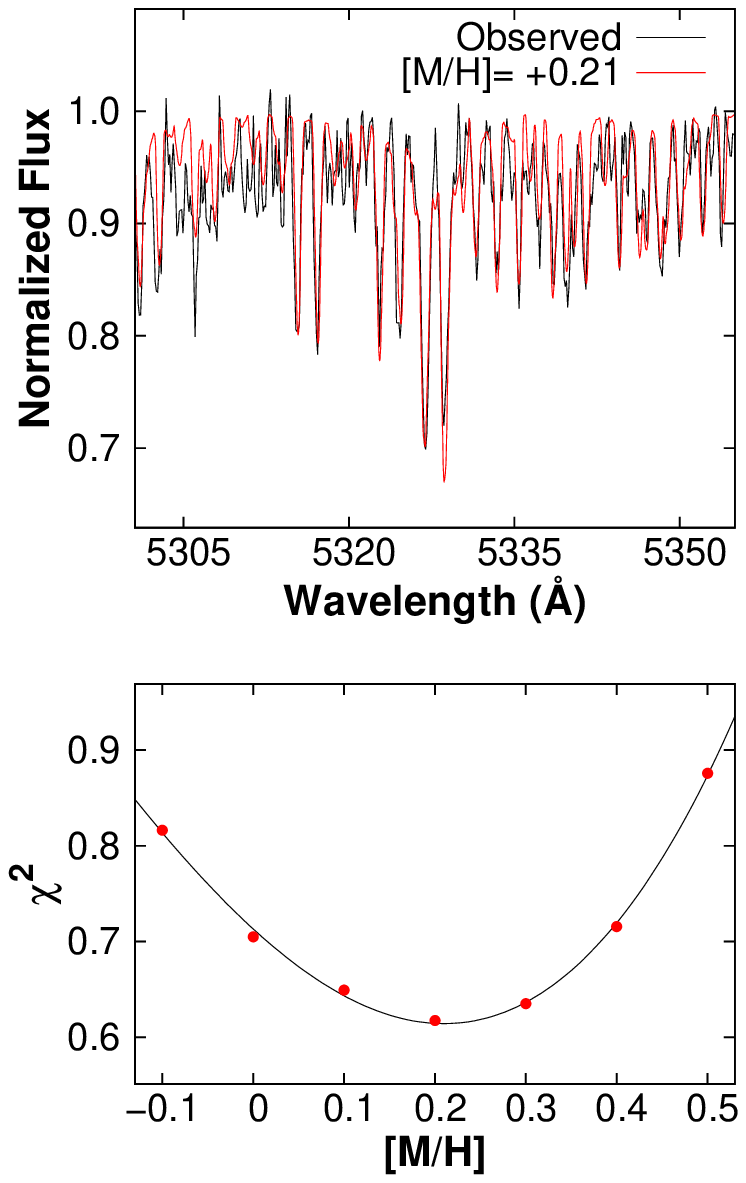}
    \end{subfigure}
    \begin{subfigure}
        \centering
        \includegraphics[width=0.3\textwidth]{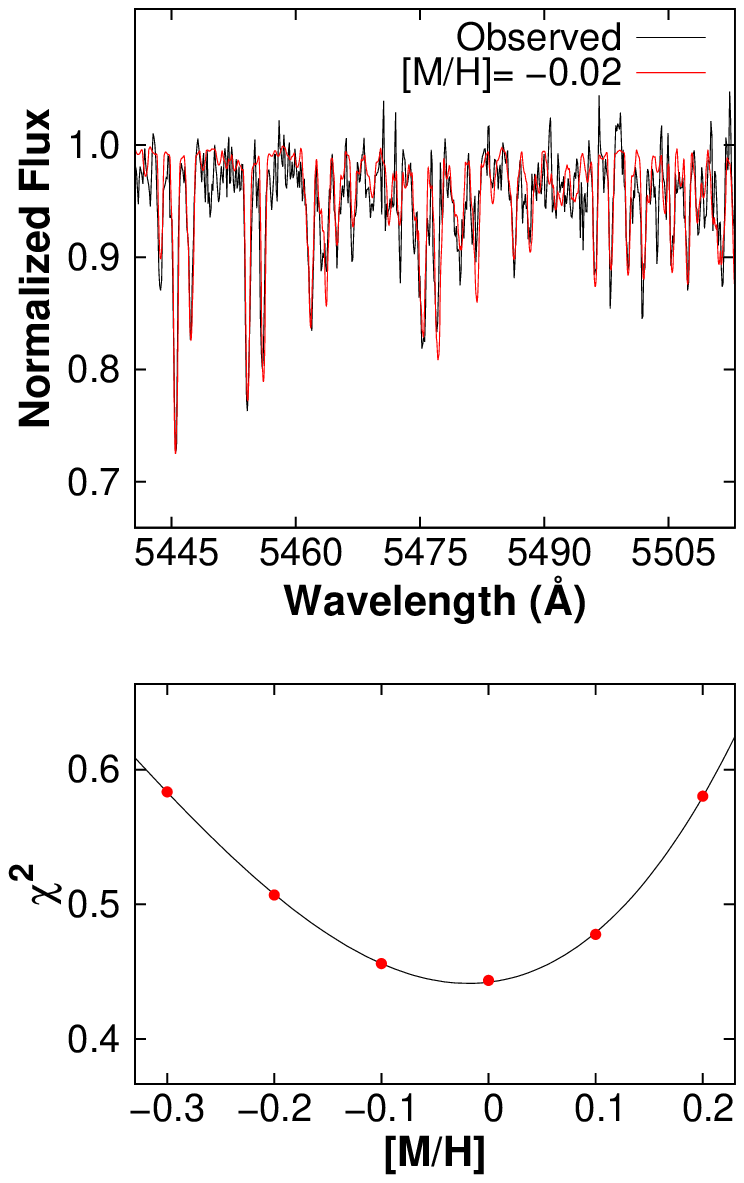}
    \end{subfigure}
    \caption{Selected regions of AN Cam for metallicity analysis}
\end{figure*}

\begin{figure*}
    \centering
    \centering
    \begin{subfigure}
        \centering
        \includegraphics[width=0.45\textwidth]{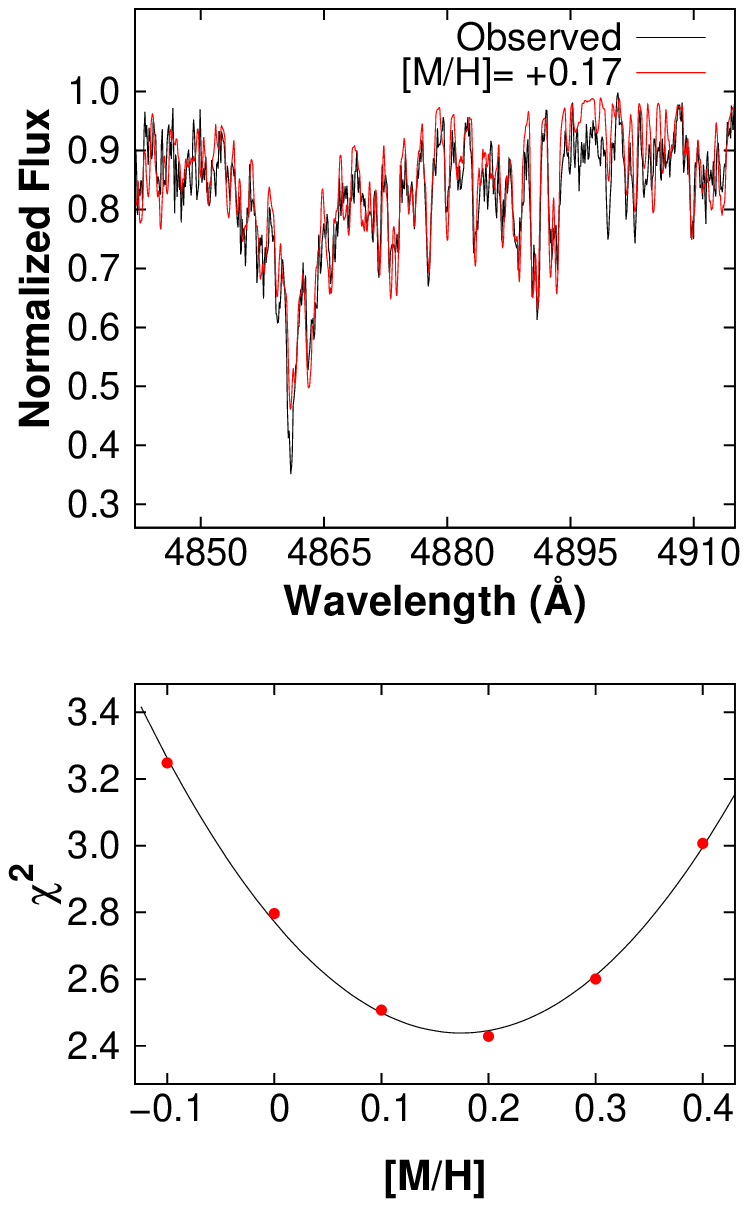}
    \end{subfigure}
    \begin{subfigure}
        \centering
        \includegraphics[width=0.45\textwidth]{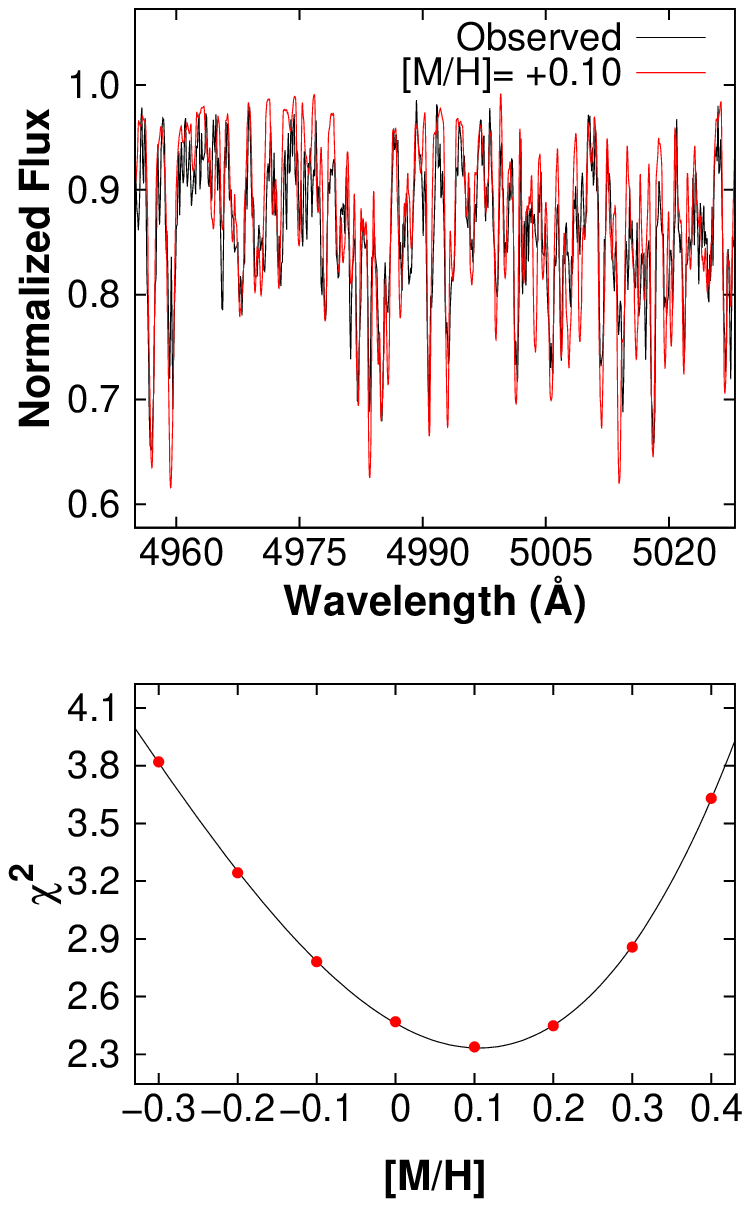}
    \end{subfigure}
    \begin{subfigure}
        \centering
        \includegraphics[width=0.3\textwidth]{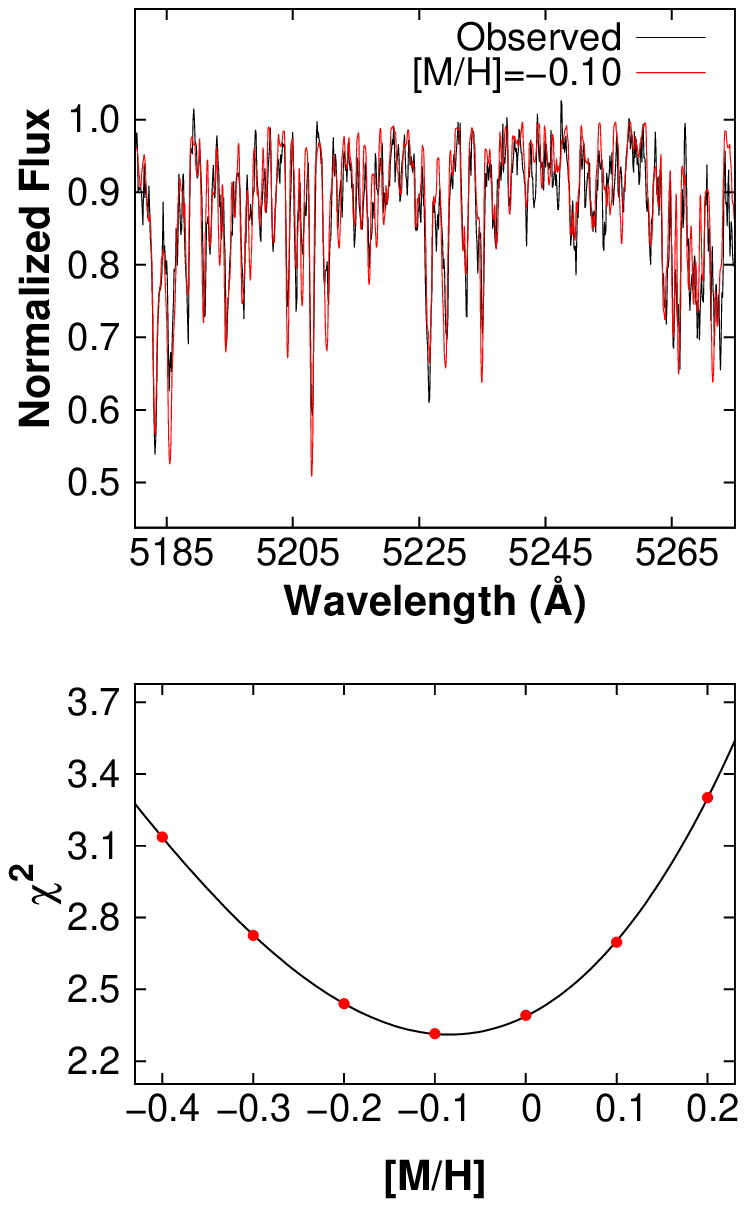}
    \end{subfigure}
    \begin{subfigure}
        \centering
        \includegraphics[width=0.3\textwidth]{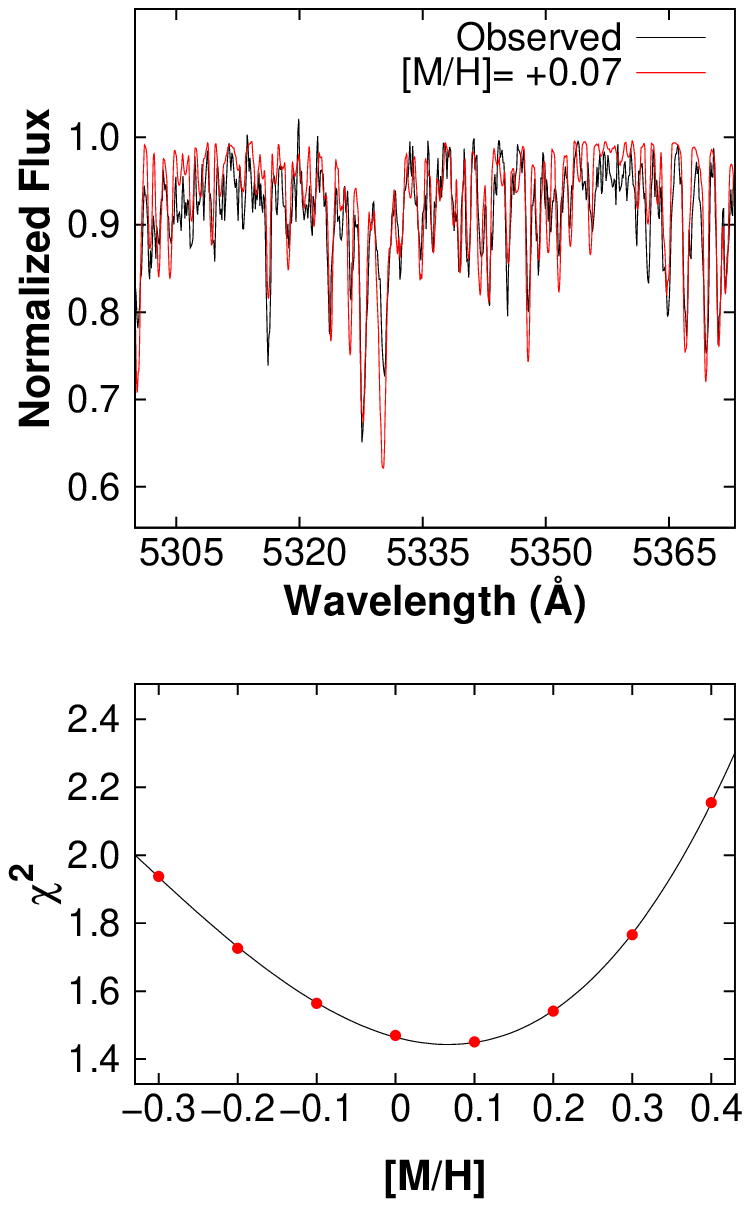}
    \end{subfigure}
    \begin{subfigure}
        \centering
        \includegraphics[width=0.3\textwidth]{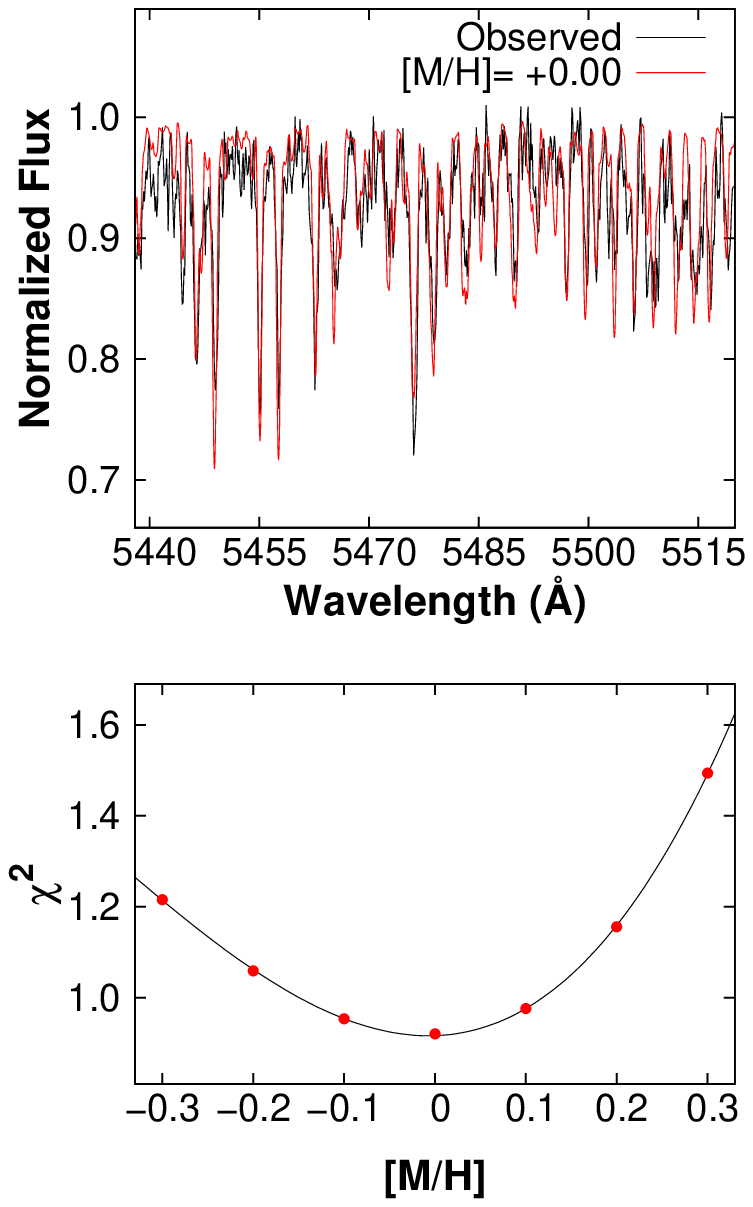}
    \end{subfigure}
    \caption{Selected regions of RS Ari for metallicity analysis}
\end{figure*}

\begin{figure*}
    \centering
    \centering
    \begin{subfigure}
        \centering
        \includegraphics[width=0.45\textwidth]{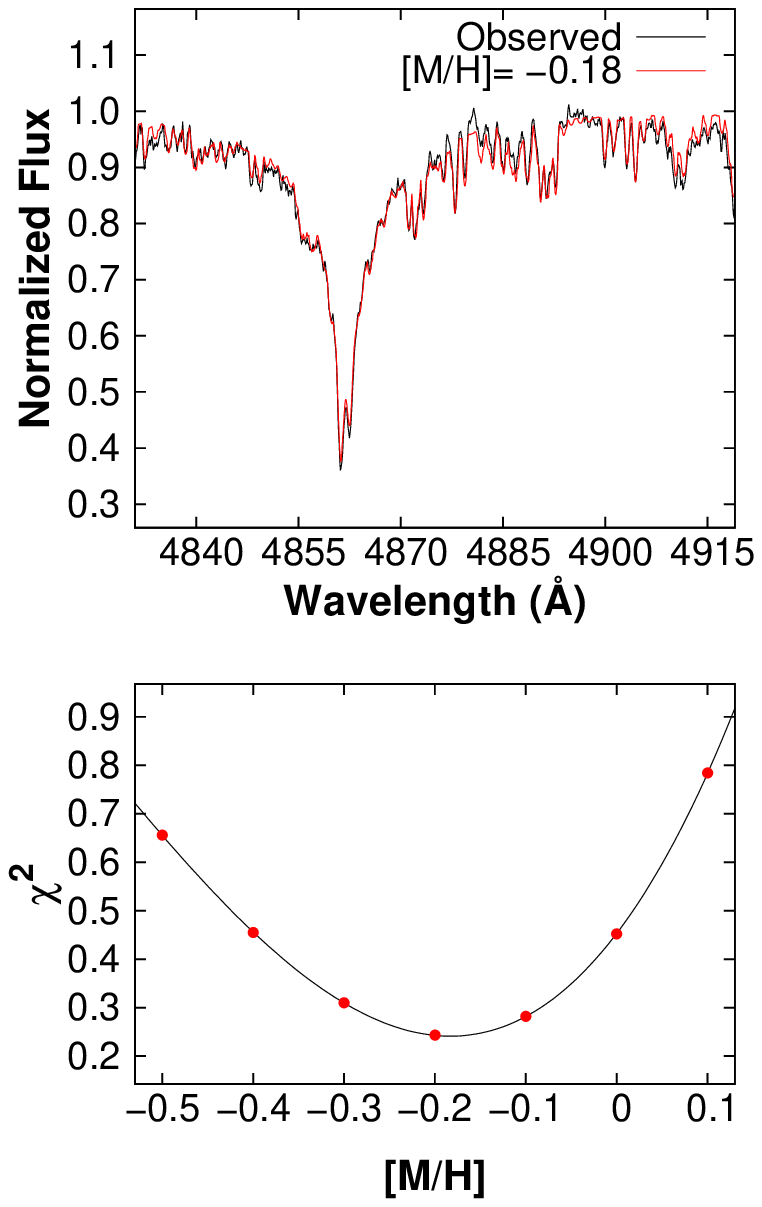}
    \end{subfigure}
    \begin{subfigure}
        \centering
        \includegraphics[width=0.45\textwidth]{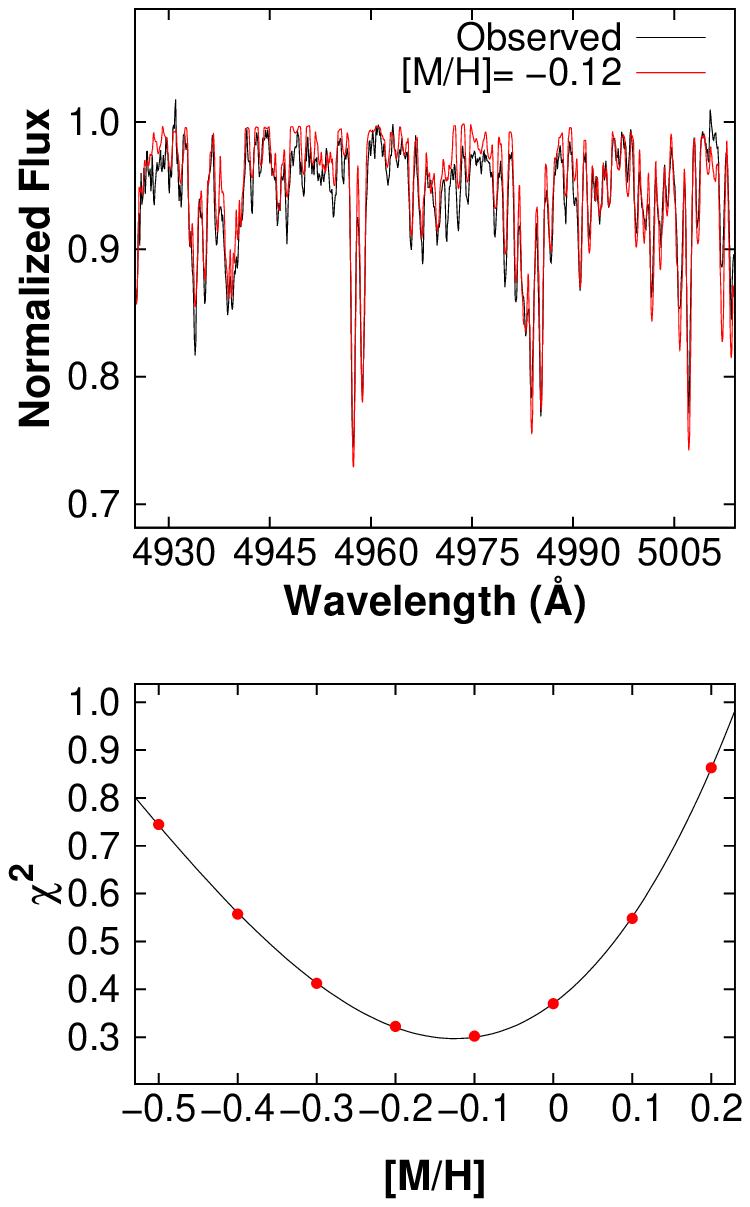}
    \end{subfigure}
    \begin{subfigure}
        \centering
        \includegraphics[width=0.3\textwidth]{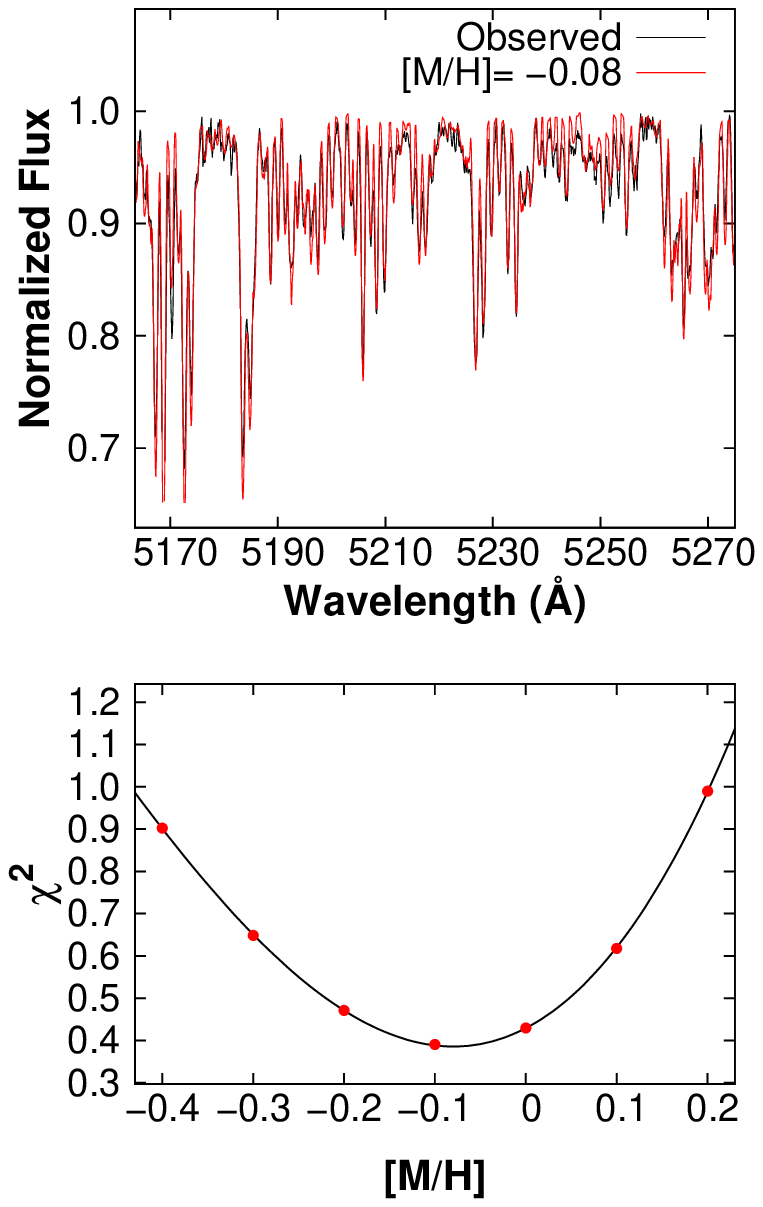}
    \end{subfigure}
    \begin{subfigure}
        \centering
        \includegraphics[width=0.3\textwidth]{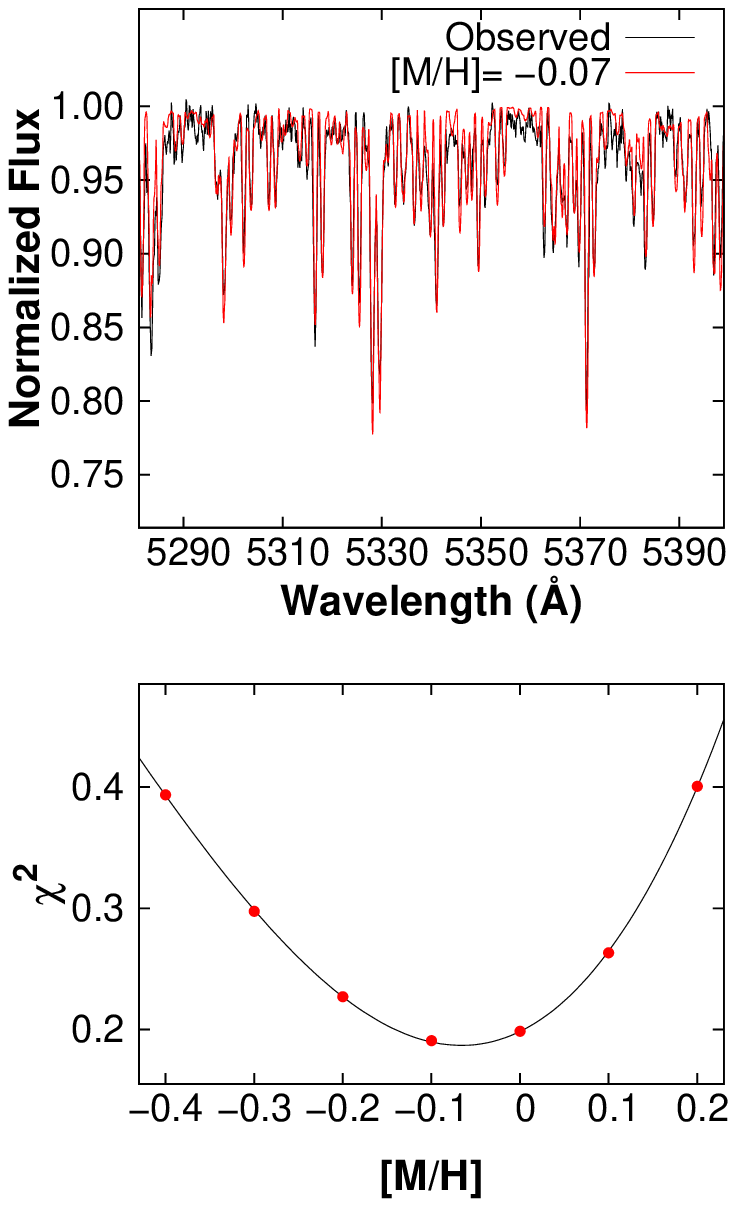}
    \end{subfigure}
    \begin{subfigure}
        \centering
        \includegraphics[width=0.3\textwidth]{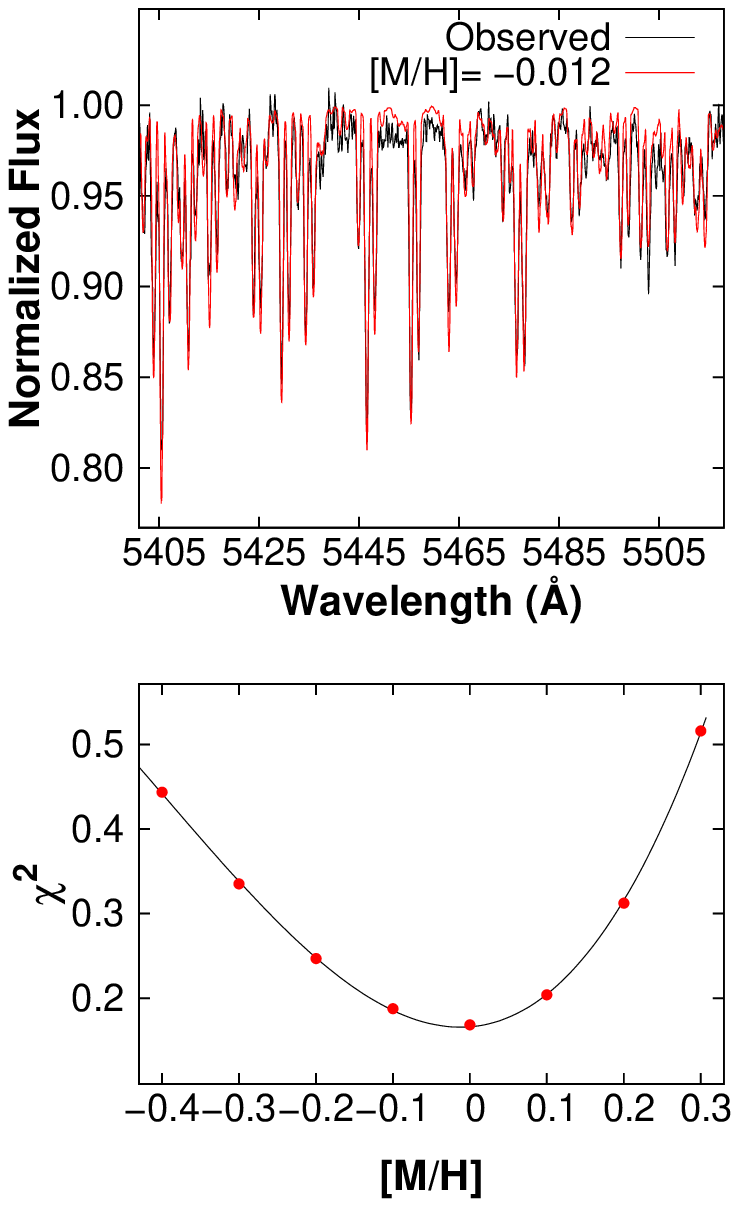}
    \end{subfigure}
    \caption{Selected regions of V455 Aur for metallicity analysis}
    \label{fig:my_v455}
\end{figure*}

\bsp	
\label{lastpage}
\end{document}